\documentclass[11pt]{article}              

\usepackage[margin=25mm]{geometry}
\usepackage{colortbl}
\usepackage{subfigure}
 
\usepackage{fancyhdr}
\usepackage{amssymb,amsfonts,amsmath,amsthm}
\usepackage{natbib}

\usepackage{authblk}
	
\usepackage{amsopn}

\newcommand{\eqdef}{\stackrel{\text{def}}{=}}

\usepackage{lipsum}
\usepackage{graphicx}
\usepackage{epstopdf}
\usepackage{algorithmic}
\ifpdf
\DeclareGraphicsExtensions{.eps,.pdf,.png,.jpg}
\else
\DeclareGraphicsExtensions{.eps}
\fi

\newcommand{\ve}[1]{\boldsymbol{#1}}			
\newcommand{\Ve}[1]{\boldsymbol{#1}}			
\newcommand{\cA}{{\mathcal A}}

\newcommand{\cU}{{\mathcal U}}

\newcommand{\cm}{{\mathcal M}}
\newcommand{\cd}{{\mathcal D}}

\newcommand{\Nn}{{\mathbb N}}

\newcommand{\ua}{\ve{\alpha}}

\newcommand{\vX}{\ve{X}}
\newcommand{\vx}{\ve{x}}

\newcommand{\cx}{\mathcal{x}}
\newcommand{\cX}{\mathcal{X}}
\newcommand{\Esp}[1]{{\mathbb E}\left[ #1 \right]}
\newcommand{\tr}{^{\textsf T}}				

\newcommand{\ie}{\textit{i.e.}}
\newcommand{\eg}{\textit{e.g.}}

\usepackage[normalem]{ulem} 
\usepackage{marginnote}
\usepackage{ifthen}

\graphicspath{{./},{./Figures/}}

\begin{document}
\title{On optimal experimental designs for Sparse Polynomial Chaos Expansions} 

\author[1]{N. Fajraoui} \author[1]{S. Marelli} \author[1]{B. Sudret} 

\affil[1]{Chair of Risk, Safety and Uncertainty Quantification,
  
  ETH Zurich, Stefano-Franscini-Platz 5, 8093 Zurich, Switzerland}


\date{}
\maketitle

\abstract{Uncertainty quantification (UQ) has received much attention in the literature in the past decade. In this context, Sparse Polynomial chaos expansions (PCE) have been shown to be among the most promising methods because of their ability to model highly complex models at relatively low computational costs. A least-square minimization technique may be used to determine the coefficients of the sparse PCE by relying on the so called experimental design (ED), \textit{i.e.} the sample points where the original computational model is evaluated. An efficient sampling strategy is then needed to generate an accurate PCE at low computational cost. This paper is concerned with the problem of identifying an optimal experimental design that maximizes the accuracy of the surrogate model over the whole input space within a given computational budget. A novel sequential adaptive strategy where the ED is enriched sequentially by capitalizing on the sparsity of the underlying metamodel is introduced. A comparative study between several state-of-the-art methods is performed on four numerical models with varying input dimensionality and computational complexity. It is shown that the optimal sequential design based on the S-value criterion yields accurate, stable and computationally efficient PCE. \\[1em] 

}

\maketitle

\section{Introduction}

\label{sec:Introduction} 

Prediction of the behavior of complex physical and engineering systems is affected by diverse types of uncertainties, including imperfect knowledge of the system parameters and their variability. These uncertainties need to be appropriately considered and their impact on the quality of model predictions needs to be quantified in a rigorous way. In this context, uncertainty quantification (UQ) is a flexible framework within which parametric uncertainties can be effectively tackled. However, UQ may become intractable in the presence of expensive-to-evaluate computational models. In this context, surrogate modelling techniques have been increasingly adopted to enable complex uncertainty quantification analyses that would otherwise be impossible. 

Amongst the available surrogate modeling techniques, polynomial chaos expansions (PCE) are one of the most promising methods for uncertainty quantification \citep{ghanem2003stochastic,xiu2005high}. In PCE, the model response is expanded in terms of a set of orthonormal multivariate polynomials that are orthogonal with respect to a suitable probability measure. They allow one to uncover the relationships between different input parameters and how they affect the model responses.  Among the available strategies to construct PCE approximations, this paper focuses on non-intrusive least square minimization and sparse-regression-based approaches, meaning that the computational model is considered as a “black-box” model. They have been extensively proven to be particularly effective in the presence of complex models \citep{Blatman2011a}. The least-square calculation of the PCE coefficients relies on the so-called \textit{experimental design} (ED), a sample from the joint distribution of the input parameters, and the corresponding model responses. A common property of least-square techniques is that their accuracy and efficiency (in terms number of required full-model evaluations) depends on how the \textit{experimental design} is sampled. In this context, different sampling strategies have been extensively investigated in the literature {\citep{giunta2003overview,goel2008pitfalls,simpson2001sampling,simpson2001metamodels}}.  This is the focus of the design of experiments (DOE) field of research.

Two  different classes of DOE sampling strategies are of particular interest in this paper. The first class involves space-filling designs that aim at filling the space of the input parameters uniformly according to some appropriate measure. Several authors have shown that space filling designs are well suited to perform metamodelling \citep{fang2005design,simpson2001metamodels}. Among such designs, Latin hypercube sampling (LHS) methods have become very popular.  An oversampling of $2 \sim 3$ times  the number of unknown coefficients is recommended to get reliable results. These designs do not satisfy any specific optimality measure.  

The second class of interest in this paper comprises optimal experimental design strategies which rely on existing prior knowledge about the surrogate model properties. They aim at maximizing the quality of a given metamodel, with respect to some optimality criteria. A number of optimality criteria have been proposed and discussed in the literature in the context of regression applications \citep{Myers2002}. Among recent examples involving optimal experimental designs in the context of metamodelling \citet{shin2016nonadaptive} proposed a simple technique to generate efficiently quasi-optimal experimental designs for ordinary least squares (OLS) regression. It is an optimization algorithm that aims at finding the design of experiments of fixed size $N$ such that the obtained results are as close as possible to those obtained with a larger sample, prior to performing model evaluations. The optimality criterion is related to the orthogonality of the columns of the information matrix. An alternative method was introduced in \citet{zein2013efficient}, where a D-optimal criterion is optimized to make the PCE construction computationally feasible. 

In  this  paper, we are interested in selecting  the best  design of experiments in order to ensure the accuracy of the  surrogate model over the whole input parameter space with a minimum number of evaluations of the computational model. Furthermore, recent literature has shown that it is more appropriate to build up the ED sequentially by iteratively enriching an initial ED, especially when model evaluations are time consuming. Recently, in metamodeling context, sequential sampling methodologies have been identified as a more efficient strategy than one-stage sampling methods, where sample points are generated all at once \citep{jin2002sequential,lin2008model}. \citet{dos2007sequential} proposed a sequential design to improve the overall accuracy of nonlinear regression metamodels for one-dimensional and two-dimensional inputs. The approach improves the accuracy of the resulting metamodel compared to space filling designs. \citet{crombecq2011efficient} performed a comparison and analysis of different space filling sequential strategies for simulation-based modelling. A review of sequential experimental designs for metamodelling in engineering can be found in \citet{jin2002sequential}.

In this paper, we propose a novel adaptive design strategy for PCE that iteratively updates the experimental design by capitalizing on the sparsity of the underlying metamodel. The idea is to generate new sample points according to an optimality criterion based on the current metamodel estimate, so as to improve its accuracy. The new sampling points are then added to the ED and the metamodel is updated. 

The outline of the paper is as follows. In Section 2, sparse polynomial chaos expansions are introduced. In Section~3, several commonly adopted sampling strategies are reviewed. In Section 4, several sequential design strategies are reviewed. In Section 5, a novel sequential approach optimal for sparse PCE is proposed. Finally, in Section 6, a comprehensive comparative study of the different sampling strategies is performed on four numerical models with varying input dimensionality and computational complexity.

\section{Sparse Polynomial Chaos Expansions}
\subsection{Polynomial Chaos Expansions}

Let us consider a physical system whose behaviour 
is represented by a computational model $\mathcal{M}$. We suppose that this model is a  function of $M$ uncertain input parameters that are represented by independent random variables $\ve{X}=\{X_1, . . . , X_M\}$ with marginal probability density functions PDFs $\{f_{X_i}(x_i), i = 1, . . . , M\}$. As a consequence, the model scalar response denoted by $Y$ is also a
random variable:
\begin{equation}
\label{eqn:1}
Y = \cm(\Ve{X}).
\end{equation}
Provided that the output random variable $Y$ has a finite variance,  it can be 
represented as a polynomial chaos expansion as follows:
\begin{equation}
\label{eqn:2}
Y = \cm(\Ve{X})  = \sum\limits_{\ua\in\mathbb{N}^M} y_{\ua} 
{\Psi}_{\ua}(\Ve{X}).
\end{equation}
where  $ y_{\ua} \in \mathbb{R} $ are the  expansion coefficients to be 
determined, $\Psi_{\ua} 
(\Ve{X})$ 
are multivariate polynomials, $\ua \in \mathbb{N}^M$ is a multi-index $\ua=\{{\alpha_1,...,\alpha_M}\}$ that  identifies  the degree of the multivariate polynomials ${\Psi}_{\ua}$ in each of the input variables $X_i$. 
The multivariate polynomials  $\Psi_{\ua} $ are assembled as the tensor product 
of univariate  polynomials orthogonal w.r.t. 
the marginal PDF of the corresponding variable, \textit{i.e:}
\begin{equation}
\label{eqn:3}
\Psi_{\ua}(\ve{x}) = \prod_{i=1}^M \phi^{(i)}_{\alpha_i} (x_i).
\end{equation}
with
\begin{equation}
\label{eqn:univariate poly orhogonality}
\left\langle\phi^{(i)}_{\alpha},\phi^{(i)}_\beta\right\rangle
\eqdef \int_{\cd_{X_i}} \phi^{(i)}_{\alpha}(x_i)\times\phi^{(i)}_\beta(x_i) 
f_{X_i}(x_i)dx_i = \delta_{\alpha\beta}.
\end{equation}
where  $\phi^{(i)}_{\alpha_i}$ is an orthogonal polynomial in the  $i $-th variable of 
degree ${\alpha_i}$ and $\delta_{\alpha\beta} = 1$ if $\alpha = \beta$ and 
$\delta_{\alpha\beta} = 0$ otherwise. Due to this construction, it follows that the 
multivariate polynomials $\Psi_{\ve{\alpha}}$ are orthonormal w.r.t. the input 
random vector $\ve{X}$. For instance, if the input random variable $\ve{X}$ are standard normal, a possible 
basis is the family of multivariate Hermite polynomials, which are orthogonal 
with respect to the Gaussian measure. Other common distributions can be used 
together with basis functions from the Askey scheme. A more general case can be 
treated through an isoprobabilistic transformation of the input random vector $\ve{X}$ into a standard random vector.  For computational purposes, the series in Eq.~\eqref{eqn:2} is truncated after 
a finite number of terms $P$, yielding the following PC approximation:
\begin{equation}
\label{eqn:4}
Y^{PCE}= \sum\limits_{\ua\in\cA} y_{\ua}.
{\Psi}_{\ua}(\Ve{X})  
\end{equation}
where $\cA $ is a finite set of multi-indices of cardinality $P$. A standard truncation scheme, which corresponds to selecting all polynomials in the $M$ input variables of total degree not exceeding $p$, is commonly used:  
\begin{equation}
\label{eqn:5}
\cA = \{ \ua \in \Nn^M \; : \; |\ua| \le p\}.
\end{equation}
The cardinality of the set $\cA $ reads:
\begin{equation}
\label{eqn:6}
\qquad \qquad  \text{card~}\cA
\equiv P = \binom{M+p}{p}.
\end{equation}

The next step is the computation of the polynomial chaos coefficients $\ve{y}=\{y_{\ua}, \ua\in \mathcal{A}\}$. Several intrusive approaches (\eg Galerkin scheme) or non-intrusive approaches (\eg projection, least square regression) \citep{SudretRESS2008b, XiuBook2010} have  been proposed in the literature. We herein focus our analysis on least-square methods, originally introduced by \citet{ Berveiller2006a,Choi2004a} under the generic name regression. In this approach, the exact expansion can be written
as the sum of a truncated series and a residual:
\begin{equation}
\label{eqn:9}
Y= \sum\limits_{\ua\in\cA} y_{\ua} 
{\Psi}_{\ua}(\Ve{X}) +\epsilon.
\end{equation}

where $\epsilon$ corresponds to the truncated terms.  \\

The term non-intrusive indicates that the  polynomial chaos coefficients are investigated over a set of input realizations $\mathcal{X} =\{ \ve{x}^{(1)},..., \ve{x}^{(N)}\} $, referred to as the experimental design (ED). One can employ Monte Carlo sampling methods, Latin Hypercube sampling (LHS, see \citet{McKay1979}), or quasi-random sequences such as the Sobol' or Halton sequence to sample points of the ED. Denoting by  $\mathcal{Y}=\{ \cm(\Ve{x}^{(1)}),..., \cm(\Ve{x}^{(N)})\} $ the set of outputs of the computational model $\mathcal{M}$ for each point in the ED $\mathcal{X}$, the set of coefficients   $ \ve{y}={ \{y_{\ua} , \ua \in \cA \}} $ is then computed by minimizing the least square residual of the polynomial approximation over the ED $\mathcal{X}$:
\begin{equation}
\label{eqn:PCE:Theory:y_alpha minimization}
\hat{\ve{y}}=\underset{\ve{y}_{\ua}\in \mathbb{R}^P} {\mathrm{\arg\!\min}} \frac{1}{N} \sum_{i=1}^{N} \left(\mathcal{M}\left(\Ve{x^{(i)}}\right)-\sum\limits_{\ua\in \mathcal{A}} y_{\ua} 
{\Psi}_{\ua}(\Ve{x^{(i)}}) \right)^2.
\end{equation}

The ordinary least-square
solution of Eq.~\eqref{eqn:PCE:Theory:y_alpha minimization} can be formulated equivalently in matrix notation as follows: 
\begin{equation}
\label{eqn:regression_minimization}
\hat{\ve{y}}=(\ve{A}\tr{}\ve{A})^{-1}\ve{A}\tr{}\mathcal{Y}. 
\end{equation}

where   \begin{center}
	$ \ve{A}=\{A_{ij}=\psi_j(\ve{x}^{(i)}) $,  $i=1,...,N$; $j=0,...,P\}. $\\
\end{center}
in which $ \ve{A} $ is the model matrix that contains the values of all the polynomial basis evaluated at the experimental design points.\\

Solving Eq.~\eqref{eqn:regression_minimization} requires a minimum number of sample points. One typically selects $N >\text{card}\cA$ to ensure the well-posedness of the regression problem. As a rule of thumb, the number of samples is often chosen such that $N \approx 2 - 3 \, \text{card}\cA$. Due to the sharp increase in the number of terms in Eq.~\eqref{eqn:4}
(see Eq.~\eqref{eqn:6}) , the required number  of model evaluations becomes prohibitive when the polynomial degree $p$ is large. 
The truncation scheme proposed in Eq.~\eqref{eqn:5} may thus lead to unaffordable calculations especially in high dimensions. This problem is known as the curse of dimensionality. As a way to mitigate this limitation, a so-called \textit{hyperbolic} truncation scheme was proposed in \citet{Blatman2010b}. It corresponds to selecting all multi-indices that satisfy:
\begin{equation}
\label{eqn:7}
\cA^{M,p,q}= \{ \ua \in \cA\; : \; ||\ua||_q \le p \},
\end{equation}
where:
\begin{equation}
\label{eqn:8}
||\ua|| = \left( \sum\limits_{i = 1}^M \alpha_i^q\right)^{1/q},
\end{equation}
and $0 < q < 1$. 
Lower values of $q$ limit the number of high order interaction terms, leading to \textit{sparse} solutions. The value $q=1$ corresponds to the standard truncation shown in Eq.~\eqref{eqn:5}.

\subsection{Sparse Polynomial Chaos Expansions}
\label{sec:SPCE}
To further sparsify the solution, without sacrificing possibly important interaction terms, \citet{Blatman2011a} proposed an adaptive sparse PCE calculation strategy based on the least angle regression (LAR) algorithm \citep{Efron2004}. Based on a given experimental design, only regressors that have the greatest impact on the model response are retained. LAR consists in directly modifying the least-square minimization problem in Eq.~\eqref{eqn:PCE:Theory:y_alpha minimization} by adding a penalty term of the form $\lambda || \ve{y}||_1$, \ie{} solving:
\begin{equation}
\label{eqn:PCE:Theory:Penalized Regression Eq}
\ve{y}= \underset{\ve{y}_{\ua} \in \mathbb{R}^P}{\arg\!\min}~
\frac{1}{N} \sum_{i=1}^{N} \left(\mathcal{M}\left(\Ve{x^{(i)}}\right)-\sum\limits_{\ua\in \mathcal{A}} y_{\ua} 
{\Psi}_{\ua}(\Ve{x^{(i)}}) \right)^2
+ \lambda||\ve{y}||_1,
\end{equation}
in which $||\ve{y}||_1 = \sum\limits_{\ve{\alpha}\in\cA}|y_{\ve{\alpha}}|$ is the regularization term that forces the minimization to favour low-rank solutions. In this study, we adopt the hybrid Least Angle Regression (LAR) method for building sparse PCE. For further details, the reader is referred to \citet{BlatmanJCP2011}. \\

\subsection{Measures of accuracy}\label{Measures of accuracy}

Once the metamodel is built, it is important to assess its quality.  A natural measure is 
provided by the  mean square error $ MSE$, defined as:
\begin{equation}
\label{eqn:MSE}
MSE = \Esp{ (\cm(\vX) - \cm^{PC}(\vX))^2} \approx 
\frac{1}{N_{val}}\sum\limits_{i = 
	1}^{N_{val}} \left( \cm(\Ve{x}^{(i)}) - 
\cm^{PC}(\Ve{x}^{(i)})\right)^2
\end{equation}
where the validation set $\{\ve{x}^{(1)},\cdots,\ve{x}^{(N_{val})}\}$ consists of samples of the 
input vector that do not belong to the experimental design and $\cm^{PC}$ is 
the surrogate model. In practice, one needs a sufficiently large validation 
set (typically in the order of $N_{val} \sim 10^{4-5}$) to achieve a stable  
estimate of the $MSE$. A low $MSE$ corresponds to an accurate metamodel. A common variant of the 
MSE error in Eq.~\eqref{eqn:MSE} is the relative mean square error $RMSE$, 
defined as: 
\begin{equation}
\label{eqn:RMSE}
RMSE = \frac{MSE}{\sigma_Y^2},
\end{equation}
where $\sigma_Y^2$ is the variance of the model response or its estimator: 

\begin{equation}
\label{variance}
\sigma_Y^2=\frac{1}{N_{val}}\sum_{i=1}^{N_{val}} \left( \cm(\Ve{x}^{(i)})-\mu_{\cm}\right)^2, \quad \mu_{\cm}=\frac{1}{N_{val}}\sum_{i=1}^{N_{val}} \cm(\Ve{x}^{(i)}).
\end{equation}\\

In pactice, this error is only affordable for analytical functions, since it requires evaluating a large number of model responses. As an alternative, an error estimate based on the ED may be used instead. In particular, the leave-one-out (LOO) error, denoted as $\epsilon_{LOO}$, has been proposed \citep{Geisser:1975,Stone1974} in the context of PCE. The LOO error is a cross-validation technique developed in statistical learning theory. It consists in rebuilding sequentially $N$ metamodels, denoted $\cm^{PC\backslash i}$, using  the original experimental design excluding one point  ${\mathcal{X}}\backslash{\Ve{x}^{(i)}} 
= \{\ve{x}^{(j)}, j=1,...,N, j\neq i\}$, and then calculating the prediction error at the withheld point $\Ve{x}^{(i)}$ (see, e.g., \citet{Blatman2011a}). The general formulation of the leave-one-out error is:
\begin{equation}
\label{eqn:PCE:Theory:LOOError}
\epsilon_{LOO} = \frac{\sum\limits_{i = 1}^N \left( \cm(\Ve{x}^{(i)}) - 
	\cm^{PC\backslash i}(\Ve{x}^{(i)})\right)^2}{\sum\limits_{i = 1}^N 
	\left(\cm(\ve{x}^{(i)}) - {\mu}_\mathcal{M}\right)^2}.
\end{equation}

When the model is a linear superimposition of orthogonal terms, as in the case for PCE, the LOO error does not require the actual calculation of $N$ independent surrogates on $N-1$ experimental design points each. Instead, it can be calculated analytically from a single surrogate based on all the $N$ points according to the following equation:
\begin{equation}
\label{eqn:PCE:Theory:ModifiedLOOError}
\epsilon_{LOO} = {\sum\limits_{i = 1}^N \left( 
	\frac{\cm(\Ve{x}^{(i)}) - 
		\cm^{PC}(\Ve{x}^{(i)})}{1-h_i}\right)^2}\bigg/{\sum\limits_{i = 1}^N 
	\left(\cm(\ve{x}^{(i)}) - {\mu}_\mathcal{M}\right)^2},
\end{equation}
where $h_i$ is the $i^{th}$ diagonal term of matrix $ \ve{A}(\ve{A}^{T}\ve{A})^{-1}\ve{A}^{T} $.\\

\section{Space-filling designs of experiments}
\label{Sampling methods}

The LAR algorithm allows one to detect automatically the significant terms in the PC expansion. However, its accuracy is greatly dependent on the sampling strategy adopted to generate the ED. Space-filling techniques are a common choice in this context. This class of designs aims at filling the input space $\cd_{\ve{X}}$ as uniformly as possible for a given number of samples $N$. In this section, we briefly review the space filling sampling techniques adopted in our study. Without loss of generality, we will consider independent uniform sampling of the unit hypercube. The more general case ( for example sampling of non-uniform distributions) can be achieved by considering isoprobabilistic transforms (e.g., Nataf or Rosenblatt transformation, \citep{rosenblatt1952remarks}).

\subsection{Monte Carlo sampling }

Monte Carlo is a popular method for generating samples from a random vector. Samples are drawn randomly according to the probability distributions of the input variables, using a random number generator. 
However, the MC sampling method is not particularly efficient. A problem of clustering arises when small sample sizes are considered. In other words, it is likely that there are regions of the parameter space that are scarcely covered by the sampling, as well as regions in which random points are clustered together. Therefore, to reliably cover the entire parameter space, a large number of simulations is often needed, hence resulting in extensive computational costs. 

\subsection{Quasi Monte Carlo sampling}
Quasi Monte Carlo sampling, also known as low discrepancy sequences (LDS), is a family of sampling 
strategies based on the deterministic placement of sample points as uniformly as possible, 
avoiding large gaps or clusters. Uniformity is quantified in terms of \textit{discrepancy} from the uniform 
distribution. The advantage of using low discrepancy sequences is in general a faster convergence rate of stochastic estimators with respect to the standard MC  method. Several sequences are available in the 
literature, \textit{e.g.} Halton \citep{halton1960efficiency}, Faure \citep{faure1982discrepance} and Sobol’ \citep{soboldistribution} sequences, all of which are based on the Van der Corput sequence. Several practical studies have proven that the Sobol' sequence outperforms the other LDS in regression 
applications \citep{kucherenko2015exploring}, therefore it is the one of choice in this paper. 

\subsection{Latin hypercube sampling}
Latin hypercube sampling is widely applied in computational engineering. It was originally introduced by \citet{McKay1979} as an improvement of the Monte Carlo sampling technique. It is a form of stratified sampling technique that improves convergence w.r.t. MCS thanks to a better representation of the sample space. 
It was shown in \citet{stein1987large} that the variance of the LHS sample mean is asymptotically (i.e. in the limit of a large number of samples) lower than traditional MCS. 
In its original formulation in \citet{McKay1979}, LHS consists in dividing the domain of each input variable 
in $N$ intervals of equal probability, followed by drawing a random sample inside each interval. A Latin hypercube design of $N$ points in $M$ dimensions can be built from a $N\times M$ matrix $\pi_{ij}$, where each column is a random permutation of the sequence $\{1, 2,..., N\}$. The design points are 
then defined as follows: 
\begin{equation}
\label{eqn:13}
\mathcal{X}_{LHS}=\{x^{(i)}_{j}=\frac{\pi_{ij}-1+u_{ij}}{N},  i=1,...,N; 
j=1,...,M\}.
\end{equation} 
where $u_{ij}$ is a realization of a uniform random variable $U \sim \cU(0,1) $. It is easy to verify that one dimensional projections of the LHS design are uniformly spaced. However, this property is not guaranteed in principle in $M$-dimensional space. Because of the random nature of LHS, some configurations exist such that the input variables are correlated, resulting in poor space filling qualities. Several modifications of the sampling scheme have been proposed in the literature leading to the so-called optimal Latin hypercube sampling designs (OLHS). One common strategy is to combine standard LHS with a space filling criterion such as entropy, maximin distance \citep{pronzato2012design}, low discrepancy or minimum correlation between the samples. In general, an OLHS is achieved by creating several independent LHS designs of the desired size, then choosing the one that best satisfies the chosen optimality criterion.\\

The main difficulty with  LHS designs is that the sampling size has to be selected \textit{a priori.} To circumvent this problem, \citet{Blatman2010b,rennen2009subset,Wang2003} developed the so-called nested-LHS designs (NLHS). The main idea is to enrich an existing LHS design such that the resulting combined design meets the LHS requirements as closely as possible. Enriching an existing LHS of size $N$ with $N_{en}$ points can be achieved by substituting the partitioning of the unit hypercube $\pi_{ij}$ in Eq.~\eqref{eqn:13} with a finer one of size $N+N_{en}$. Then, $N_{en}$ empty partitions are identified and a random sample is drawn in each of them. While this procedure does not guarantee that the final design is still an LHS (due to the possibility that multiple points of the original design lie in the same stratum after the new partitioning), the properties of the final design tend to be closer to those of an LHS. For more details, see \citet{Blatman2010b,Wang2003}.

\subsection{Measures of quality for space filling designs}
\label{sec:measures of quality}

A number of different criteria for assessing the space filling quality of 
a particular experimental design can be found in the literature. In the 
following, we briefly recall two widely used space-filling criteria, allowing 
for enhanced discrimination when choosing among candidate designs. The first 
one is the maximin distance criterion \citep{johnson1990minimax}, which 
aims at maximizing the smallest distance between neighboring points, 
\textit{i.e.}:   
\begin{equation}
\label{eqn:14}
\text{maximin}(\mathcal{X}) \eqdef
\underset{\ve{x} \in \mathcal{X}} \max \, \underset{\ve{y} \in 
	\mathcal{X}}\min \,  d(\ve{x}, \ve{y})
\end{equation} 
where $d(\ve{x}, \ve{y})$ is the Euclidean distance between two sample points $ \ve{x}$ and $\ve{y}$. A second criterion was proposed to assess the uniformity quality of the design based on a discrepancy measures. To this purpose, a common and computationally efficient tool is the centred $D^2$ discrepancy criterion, which reads:
\begin{equation}
\label{eqn:15}
\begin{split}
D^2({\mathcal{X}})= &\left(\frac{13}{12}\right)^M-\frac{2}{N} \sum \limits_{i 
	= 	1}^N \prod \limits_{k = 1}^M \left(1+\frac{1}{2} \lvert 
x^{(i)}_k-\frac{1}{2} \rvert-\frac{1}{2} \lvert x^{(i)}_k-\frac{1}{2} 
\rvert^2\right)+\\
&\dfrac{1}{N^2} \sum \limits_{i,j = 1}^N \prod \limits_{k=1}^M \left(1+\frac{1}{2} \lvert x^{(i)}_k-\frac{1}{2} \rvert-\frac{1}{2} \lvert x^{(j)}_k-\frac{1}{2} \rvert-\frac{1}{2} \lvert x^{(i)}_k-x^{(j)}_k \rvert\right)
\end{split}
\end{equation} 
A design is called uniform if it shows a low centered $D^2$ discrepancy. 

\section{ Optimal designs of experiments }
\label{sec:optimal sequential designs}

Optimal designs leverage on the knowledge about the properties of the surrogate model that is being considered. This can be achieved by searching a design $\cX$ that optimizes a given optimality criterion. For instance, in the context of regression-based surrogate models (\eg{} PCE), \citet{zein2013efficient} proposed a sampling method for regression-based polynomial chaos expansion that consists in maximizing the determinant of the information matrix $A\tr{}A$ in Eq.~\eqref{eqn:regression_minimization} for a given number of samples, by using the D-optimality criterion. Recently, \citet{shin2016nonadaptive} proposed a point selection approach to construct an optimal experimental design with the aim of providing an accurate OLS estimation. The optimal design is determined by maximizing a quantity that combines both the determinant and the column orthogonality of the information matrix, denoted $S$-value criterion. In the following section, we will briefly review the D-optimality criterion and $S$-value criterion  for regression-based PCE. 

\subsection{D-optimal designs }
\label{sec:D-optimal}
The D-optimality criterion is aimed at minimizing the variance of the regression parameters estimate. The latter is derived from Eq.~\eqref{eqn:regression_minimization} as follows:
\begin{equation}
\label{eqn:varianceD}
\begin{split}
\text{Var}(\hat{\ve{y}}) = & \text{Var}((\ve{A}\tr{}\ve{A})^{-1}\ve{A}\tr{}\mathcal{Y} )\\
= & (\ve{A}\tr{}\ve{A})^{-1}\ve{A}\tr{}\text{Var}(\mathcal{Y} )    ((\ve{A}\tr{}\ve{A})^{-1}\ve{A})\tr{}.
\end{split}
\end{equation}
Under the assumption of independent normal model errors with constant variance ($\text{Var}(\mathcal{Y})=\sigma^2\ve{I}$), we obtain:  
\begin{equation}
\label{eqn:17}
{\text{Var}(\hat{\ve{y}})=(\ve{A}\tr \ve{A})^{-1}\sigma^2}.
\end{equation} 

Reducing the variance of the PCE coefficients leads to the minimization of the determinant $\text{det} (\ve{A}\tr{}\ve{A})^{-1}$. To avoid the inversion, one maximizes the determinant of the information matrix $\text{det} (\ve{A}\tr{}\ve{A})^{-1}$. The $D$-optimal design, denoted $\mathcal{X}_D$, is then the solution of the following optimization problem over the input design space $\cd_{\ve{X}}$: 
\begin{equation}
\label{eqn:18}
\mathcal{X}_D= \underset{\cX \in \cd_{\ve{X}}}{\arg\,\max}\, \text{det} (\ve{A}\tr{}\ve{A}).
\end{equation} 

\subsection{S-optimal designs }
\label{sec:S-optimal}

We herein briefly review the optimal approach developed in \citet{shin2016nonadaptive}. The goal is to find the optimal 
$N$-subset within a large-size sample $\mathcal{X}_{L}$ of size $N_{L}$ (typically in the order of $N_L = 10^{5-6}$) that provides the most accurate PCE. It is a subset selection problem where potential solutions are viewed as a $N\times M$ matrix $\mathcal{X}$ consisting of exactly $N$ different rows of $\mathcal{X}_{L}$. The minimization problem in Eq. ~\eqref{eqn:PCE:Theory:y_alpha minimization} over the ED $\mathcal{X}$ reads:

\begin{equation}
\label{eqn:minimization1}
\hat{\ve{y}}=\underset{\ve{y}\in \mathbb{R}^{\text{card}\cA}} {\mathrm{\arg\!\min}} \frac{1}{N} \sum_{i=1}^{N} \left(\mathcal{M}\left(\Ve{x^{(i)}}\right)-\sum\limits_{\ua\in \mathcal{A}} y_{\ua} 
{\Psi}_{\ua}(\Ve{x^{(i)}}) \right)^2.
\end{equation}

Equivalently, the minimization problem in Eq. ~\eqref{eqn:PCE:Theory:y_alpha minimization} over the large ED, $\mathcal{X}_{L}$, reads: 

\begin{equation}
\label{eqn:minimization2}
\hat{\ve{y}}_L=\underset{\ve{y}\in \mathbb{R}^{\text{card}\cA}} {\mathrm{\arg\!\min}} \frac{1}{N_L} \sum_{i=1}^{N_L} \left(\mathcal{M}\left(\Ve{x_L^{(i)}}\right)-\sum\limits_{\ua\in \mathcal{A}} y_{\ua} 
{\Psi}_{\ua}(\Ve{x_L^{(i)}}) \right)^2.
\end{equation}

The aim is to find a subset such that the estimated PCE coefficients based on the optimal design are as close as possible to those based on the large ED, $\mathcal{X}_{L}$. This is mathematically equivalent to the following optimization problem: 
\begin{equation}
\label{eqn:21}
\mathcal{X}^{\dagger}=\underset{\mathcal{X}\subset \mathcal{X}_L} {\mathrm{\arg\!\min}} (\| \hat{\ve{y}} \left(\mathcal{X}\right) - \hat{\ve{y}_L} \|).
\end{equation}
This can be intuitively interpreted as a row selection problem, such that:    
\begin{equation}
\label{eqn:22}
R^{\dagger}=\underset{R} {\mathrm{\arg\!\min}} (\| \hat{\ve{y}}\left(R\mathcal{X}_L\right) - \hat{\ve{y}_L} \|),
\end{equation}
where $R=[R_{(1)};...;R_{(N)}] \in \mathbb{R}^{N\times N_L}$ denotes a row selection matrix, where each row $R_{(i)}=[0,\cdots,1,0,\cdots,0]$ is a vector of length-$N_L$  with all zeros but a one at $k_i$ location, $1 \leq k_i \leq N_L$. As is the case in many subset selection algorithms, QR decomposition is used to find the most representative rows. Let $\ve{A}_L$ denote the model matrix obtained when considering $\mathcal{X}_{L}$ and the corresponding model evaluations  $\mathcal{Y}_L$  . let $  \ve{A}_L=Q_LR_L  $  be the $ QR$ factorization of $\ve{A}_L$. It is then possible to prove that: 
\begin{equation}
\label{eqn:23}
\| \hat{\ve{y}} - \hat{\ve{y}_L} \| \leq \| R^{-1}_L  \|\cdot \| g \|,
\end{equation}
where $ \hat{g} $ is the least square solution to 
\begin{equation}
\label{eqn:24}
RQ_Lg=R\mathcal{P^{\bot}}\mathcal{Y}_L,
\end{equation}
where $ {P^{\bot}=(I-Q_LQ_L^T)} $ is the projection operator to the orthogonal complement of the column space of $Q_L$.\\

It is then clear that achieving a smaller $\|g\|$ leads to the minimization of the difference $ \| \hat{\ve{y}} - \hat{\ve{y}_L} \|$. To do this, one seeks to: (i) maximize the determinant of $RQ_L$ so that the variance of $g$ is minimized and (ii) maximize the column orthogonality of $RQ_L$, so that the correlation between design points is minimized. For this purpose, a selection criterion combining these conditions was proposed in \citet{shin2016nonadaptive}. Such criterion is hereafter called $S$-value. For the sake of simplicity, the $S$-value will be defined for a matrix $\ve{A}$ of size $N\times \text{card~}(\cA) $ as follows:

\begin{equation}
\label{eqn:25}
\mathcal{S}\left(\ve{A}\right)=\left( \frac{\sqrt{\det \ve{A}\tr \ve{A}}}{\prod\limits_{i = 1}^{\text{card}\cA} \|\ve{A}^{(i)}\|}  \right)^{\frac{1}{{\text{card}\cA}}},
\end{equation}
where $A^{(i)}$ is the $i$-th column of $\ve{A}$. 
The optimal sample is thus the solution of the following optimization problem: 
\begin{equation}
\label{eqn:35}
R^{\dagger}=\underset{R} {\mathrm{\arg\!\max}} (\mathcal{S}(RQ_L)).
\end{equation}
It is worth noting that in the context of PCE, the columns of the model matrix $\ve{A}$ consist of orthonormal basis elements. The $QR$ decomposition is then not necessary and the matrix $\ve{A}_{L}$ can be used instead of $Q_{L}$. Thus, we seek:  
\begin{equation}
\label{eqn:50}
R^{\dagger}=\underset{R} {\mathrm{\arg\!\max}} (\mathcal{S}(R\ve{A}_L)). 
\end{equation}
The optimality condition Eq.~\eqref{eqn:50} can then be equivalently rewritten as follows: 
\begin{equation}
\label{eqn:26}
\mathcal{X}^{\dagger}= \underset{\cX \in \mathbb{X}}{\arg\,\max} (\mathcal{S}(\ve{A}\left(\mathcal{X}\right))).
\end{equation}

In their recent study, \citet{shin2016nonadaptive} considered a standard truncation scheme to select the set of basis elements. Although their optimal design has proven powerful, it may still face challenges in high dimensional cases. As already mentioned, one can reduce the size of the candidate set by considering other truncation strategies, \textit{e.g.} the hyperbolic truncation. However, it may not be known in advance which candidate set will perform best. This motivates the need of an adaptive strategy that selects iteratively the optimal candidate basis and thus the associated optimal design. 

\section{Sequential experimental designs}

In the presence of expensive computational models, it can be preferable to 
adopt a greedy iterative strategy in the construction of an experimental design 
for surrogate modelling. The class of so-called sequential experimental designs is 
based on constructing a relatively small initial space-filling design, training 
a surrogate model and assessing its accuracy. If needed, the initial design can 
then be enriched sequentially until some accuracy measure of the surrogate 
model (e.g. the number of desired points or the $RMSE$ in Eq.~\eqref{eqn:RMSE}) is satisfied.
In this paper we explore two different strategies to address this problem. The first is to enrich the initial 
experimental design according to a space-filling criterion without 
assuming any knowledge on the functional form of the surrogate model. 
The second assumes some degree of knowledge on the behaviour of the surrogate model. Note that, in any case, none of the criteria used in this study depend on the model responses on the experimental design points. \\ 
In the following, the initial experimental design $\mathcal{X}_{init}$ of size $N_{init}$ is enriched at each iteration with $N_{new}$ data points. The new points are to be selected, according to a specific criterion, from an available ED $\mathcal{X}_L$ of size $N_L$ such that $N_L>>N_{init}$. 

\subsection{Sequential selection using space filling criteria}
In this section we review a sequential approach for enriching the initial experimental design according to a maximin distance criterion. This strategy was proposed by \citet{kersaudy2015modelisation}. Note that such idea was originally developed by \citet{ravi1991facility} as a method for the dispersion problem and applied later by \citet{rennen2009subset} to subset selection for Kriging metamodels, thereby referred to as greedy maximin selection. At each iteration, the algorithm selects the point $\ve{x}^{*}\in \mathcal{X}_L$ that maximizes the minimum distance with all the points in the current experimental design $\cX$. The obtained experimental design $\cX$ of size  $N=N_{init}+N_{new}$ is therefore generated according to the following algorithm: 

\begin{enumerate}
	\item Let $\cX = \cX_{init}$ (note that $N_{init} < N$)
	\item find $\ve{x}^{*} = \underset{\vx_L \in	\cx_L}{\max}\,\underset{\ve{x}\in\cx}{\min} \left(\| \vx - \vx_L\|\right)$
	\item update $\cX = \cX \cup \vx^{*}$ and $\cX_L = \cX_L \setminus 
	\vx^*$
	\item repeat step ii until the set $\cx$ contains the desired 
	number 	of points $N$.
\end{enumerate}

The main advantage of this method is that it is easy to implement. However, the main disadvantage lies in that one needs to compute all the distances between every point of $\vx_L^{(i)}\in\cX_L$ and every point $\vx^{(j)} \in \cX$ , which can be challenging in high dimension. In the following, the obtained design is referred to as sequential maximin design. A similar idea has also been proposed in \citet{iooss2010numerical} where the centred discrepancy measure $ D^2 $ in Eq.~\eqref{eqn:15} is used instead of the maximin criterion. In this case, the algorithm enriches the initial experimental design $\cX_{init}$ with points that minimize its $D^2$ discrepancy. It operates in the same way as the greedy maximin algorithm. In this paper, we choose to limit our study to the maximin criterion based on Euclidian distance.

\subsection{Adaptive selection of the optimal design}
\label{sec:adaptive optimal designs}
The optimal design strategies presented until now rely on the assumption that the regressors in an OLS problem are known. However, in the context of sparse PCE, the optimal polynomial basis is in general not known a priori. To address this problem, we propose an iterative strategy in which the estimated sparse PCE basis is automatically updated and the experimental design is enriched. To construct a sparse PCE, we propose the use of a basis selection method based on compressive sensing techniques. In particular, we rely herein on the least angle regression algorithm (LAR) \citep{Efron2004}. Other alternatives (\eg orthogonal matching pursuit (OMP) \citep{Jakeman2015,mallat1993matching} or stepwise regression \citep{hesterberg2008least}) are also directly compatible with our procedure.

\subsubsection{Sparse PCE: Least Angle regression}
The LAR algorithm is an iterative regression method applied in the context of PCE for basis selection \citep{Blatman2011a}. The basic idea of LAR is to select a parsimonious number of regressors that have the greatest impact on the model response, among a possibly large set of candidates. LAR eventually provides a sparse polynomial chaos expansion, \ie one that only contains a small number of regressors compared to a classical full representation. 
It consists on iteratively moving regressors from a \textit{candidate set} to an \textit{active set}. The next regressor is chosen based on its correlation with the current residual. At each 
iteration, analytical relations are used to identify the best set of regression 
coefficients for that particular active set, by imposing that every active 
regressor is equicorrelated with the current residual. At the end of the process, the optimal sparse PC basis is chosen according to a suitable measure of accuracy. In particular, the leave-one-out (LOO) error, $\epsilon_{LOO}$, is a natural candidate in the context of PCE. \\

The LAR algorithm in the context of PCE \citep{Blatman2011a} can be summarized as follows:

\begin{enumerate}
	\item  Initialize the set of candidate regressors to the full basis and the active set of regressors to $\varnothing$.
	
	\item  Initialize all coefficients equal to $0$. Set the residual equal to the output vector.
	\item Find the regressor ${\Psi_{\ua_j}}$ that is most correlated with the 
	current residual
	\item Move all the coefficients of the current active set towards their 
	least-square value until their regressors are equicorrelated to the residual as 
	some other regressor in the candidate set. This regressor will also be the most 
	correlated to the residual in the next iteration. 
	\item Calculate and store the accuracy estimates, $\epsilon_{LOO}^j$, for the current iteration
	\item Update all the active coefficients and move $\ve{\Psi_{\ua_j}}$ from the 
	candidate set to the active set
	\item Repeat the previous steps (iii-v) until the size of the active set is equal to 
	$m = \min(P,N-1)$
	\item Choose the best iteration based on their prescribed error estimate, $\epsilon_{LOO}$.
\end{enumerate}

Note that after detecting the relevant basis by LAR, the coefficients are recomputed using the ordinary least-square method (hybrid-LAR). Readers are referred to \citet{Blatman2010b} for further details.

\subsubsection{Adaptive selection of the optimal design}
The algorithm proposed in the last subsection allows one to detect the most relevant regressors in the PC expansion on the basis of the current experimental design. However, it is based on an arbitrarily fixed experimental design. To address this problem, we propose an adaptive strategy that updates iteratively both the experimental design and the estimated sparse PCE basis. The initial ED generated using a space filling techniques will be enriched sequentially according to an optimality criterion. At each iteration, the optimality criterion is reformulated based on the updated sparse basis. The proposed algorithms works as follows:  

\begin{enumerate}
	\item \textbf{Initial experimental design}: we start with a small sample 
	set $\mathcal{X}_{init}$ of a size $N_{init}$, normally produced by LHS or Quasi Monte Carlo sequences. 
	\item \textbf{Basis selection}: a degree-adaptive sparse PCE \citep{BlatmanJCP2011} is then constructed using LARS to identify the sparse set of polynomials (out of a candidate set) that best approximates the computational model based on the current experimental design.
	\item \textbf{Formulation of the model matrix}: This step is the core of 
	the sequential process. The information matrix is updated according to the basis 
	obtained in the previous step. The optimality criterion ($D$-optimal or $S$-optimal, see Sections \ref{sec:D-optimal} and \ref{sec:S-optimal}, respectively) is then re-evaluated on $\mathcal{X}_L \setminus \mathcal{X}$ to identify the next set of candidate points $X_{new}$. 
	\item \textbf{Experimental design augmentation}: The identified 
	$N_{new}$ $S$- or $D$- optimal data points are added to the current 
	experimental design, in which the computational model is evaluated. The sparse PCE is then updated. 
	
\end{enumerate}

The procedure (steps ii, iii and iv) is repeated until a stopping criterion is fulfilled. In particular, the accuracy of the metamodel (estimated \eg with cross-validation techniques) or the maximum number of sample points can be used. For easier comparisons between the methods, we choose to fix the maximum number of sample points as the stopping criterion in the following.\\

The effectiveness of the adaptive methods is dependent on the optimization algorithm that searches the optimal sample points. Various search strategies have been proposed in the literature. In this work, the $D$-optimal experimental design is generated using the built-in \textsc{Matlab} function $candexch$ from the statistics toolbox. The $candexch$ routine implements the Fedorov exchange algorithm \citep{miller1994algorithm}. The $S$-optimal experimental design is generated using the greedy algorithm proposed in \citet{shin2016nonadaptive}. For a comprehensive overview of the implementation of the greedy algorithm, the reader is referred to \citet{shin2016nonadaptive}.

\section{Results and Discussion}

In this section, we investigate the efficiency of the proposed adaptive sequential sampling methods, when either the $D$-optimality or the $S$-value criteria are used to generate the sequential samples. For the sake of clarity, the sequential sampling corresponding to these two sampling settings are denoted as $Seq$ D-optimal and $Seq$ S-optimal, respectively. A comparative study with other commonly used sampling methods is performed. Four different sampling techniques are considered:  $maximin$ LHS, $min-corr$ LHS, Sobol' sequences and Sequential $maximin$ design, denoted $Seq$ maximin. $Maxmin$ LHS is obtained by selecting an LHS design from a set of five different designs according to the maximin distance criterion, while $min-corr$ LHS is obtained similarly, but with a minimum-correlation 
based selection criterion. For this purpose, we use the \textsc{Matlab} function $lhsdesign$. 
As already mentioned, D-optimal designs are generated with the \textsc{Matlab} built-in $candexch$ function. For comparison of the space-filling properties of the sample points, the measures described in Section \ref{sec:measures of quality} are used. In addition, the condition number (CN) of the information matrix is considered, as a measure of orthogonality.\\
In this section, the proposed  adaptive sampling methods are validated on four models with varying input 
dimensionality and complexity. They comprise two analytical functions, namely the Ishigami function and the Sobol' g-function, and two finite element models, namely a truss structure and a 1D diffusion problem. Because all of the presented benchmarks are inexpensive to evaluate, the relative generalization error (RMSE) on a validation set is then employed to assess the accuracy of the resulting polynomial chaos expansions.  All of the PCE metamodels, as well as the LARS basis selection, are developed within the \textsc{UQLab} software framework \citep{Marelli2014,UQdoc_09_104}. \\

Since the performance of the ED is sensitive to the random nature of the sampling methods, each analysis is replicated 50 times. To generate a randomized Sobol' sequences, several parallel sequences are generated using different starting points. The results are then summarized in box-plots. These replications aim at assessing the effect of stochastic variations in the generation of the designs. 
\subsection{Ishigami function}
The Ishigami function \citep{Ishigami90} is a well-known benchmark for polynomial chaos expansions. It is a 3-dimensional analytical function characterized by non-monotonicity and high non-linearity, given by the 
following equation:
\begin{equation}
\label{eqn:27}
f(\ve{x}) = \sin(x_1) + a \sin^2(x_2) + b x_3^4\sin(x_1).
\end{equation}
The input random vector consists of three \textit{i.i.d.} uniform random variables ${X_i \sim \mathcal{U} (-\pi, \pi)}$. The numerical values $\{a = 7,b = 0.1\}$ are chosen for this example. Note that this model is sparse in nature. \\

\begin{figure}
	\centering
	\includegraphics[width=.4\textwidth,clip = true, trim = 0 0 0 	
	0]{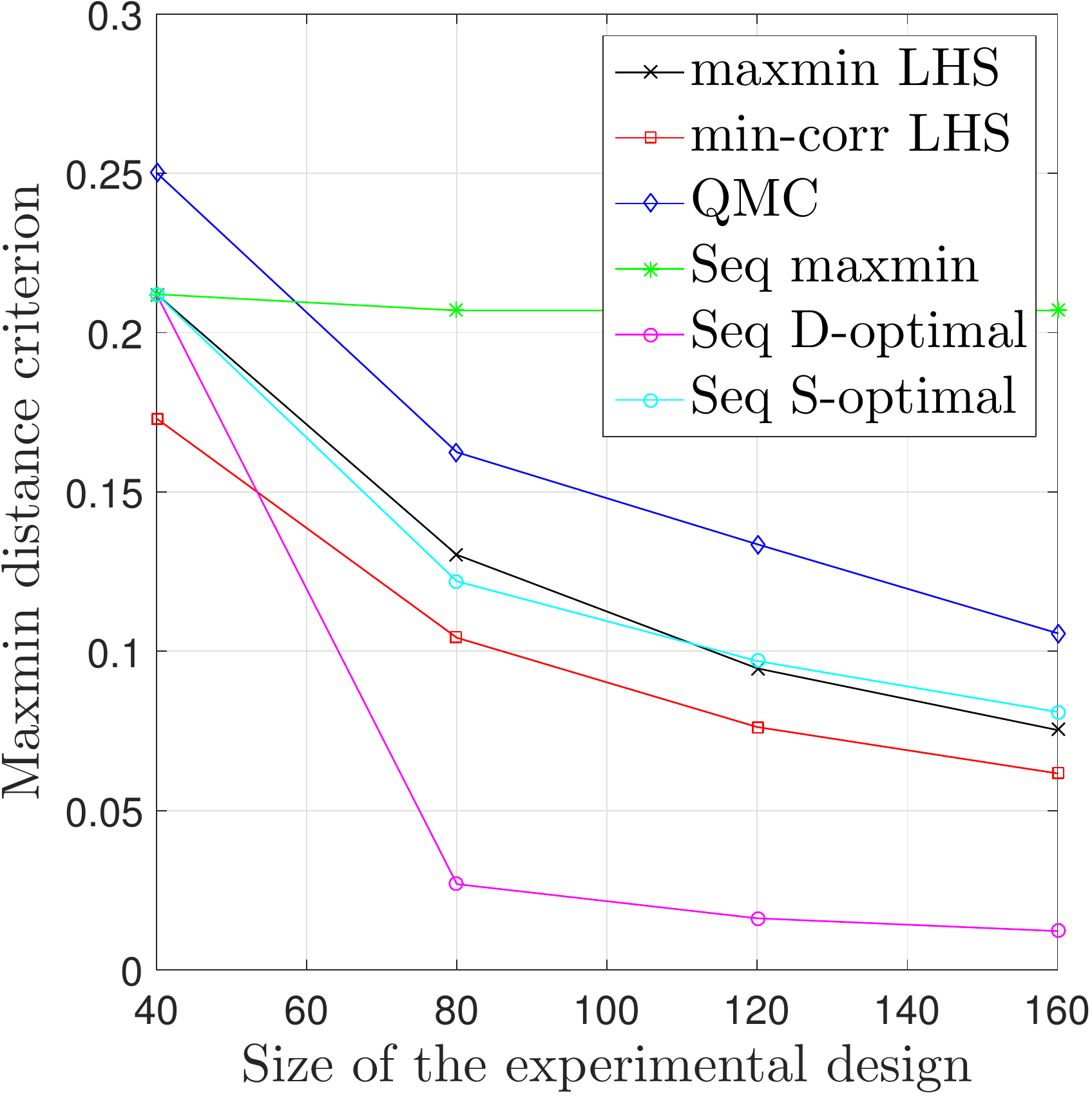}
	\includegraphics[width=.4\textwidth,clip = true, trim = 0 0 0 
	0]{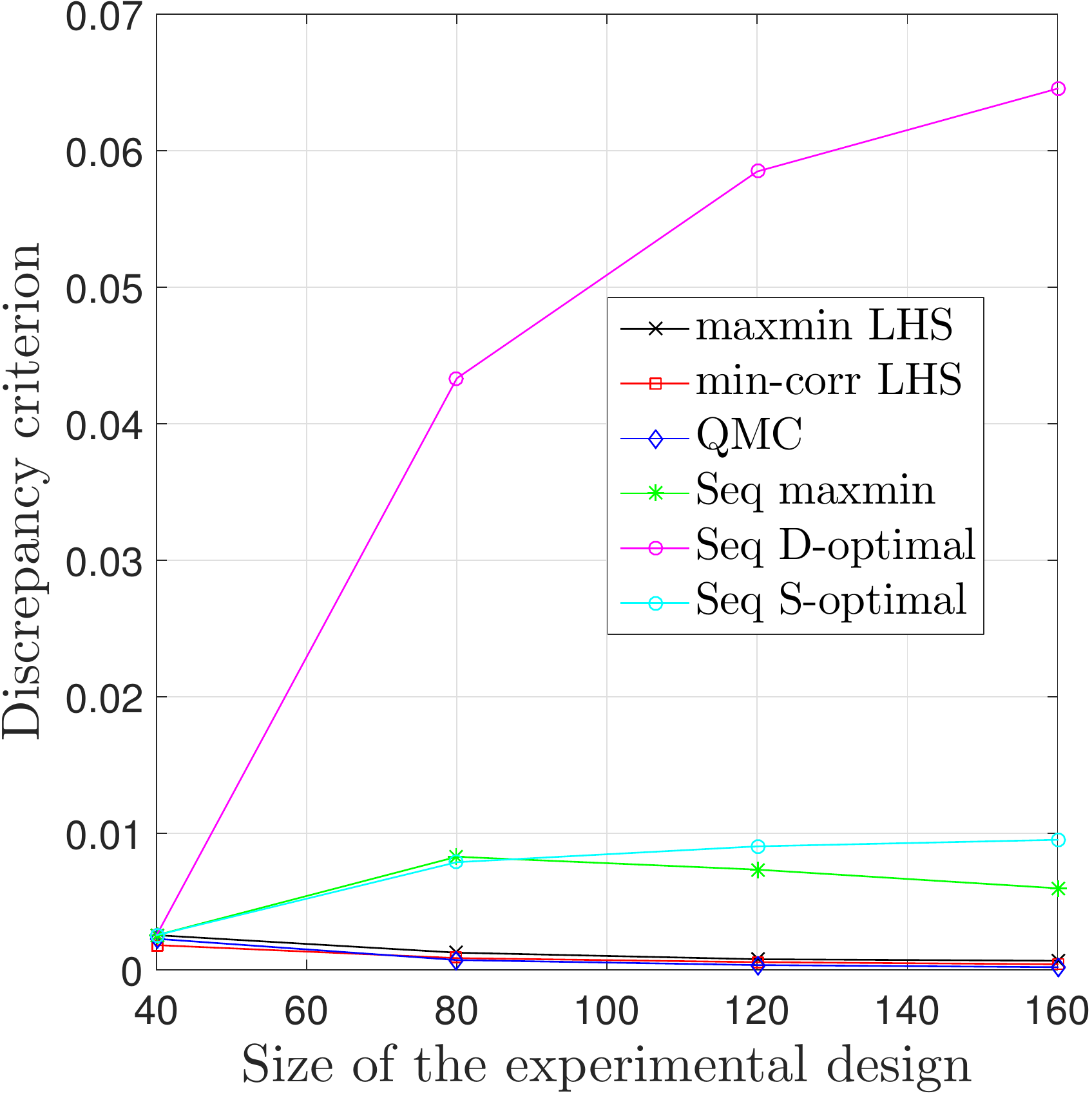}
	
	\includegraphics[width=.4\textwidth,clip = true, trim = 0 0 0 
	0]{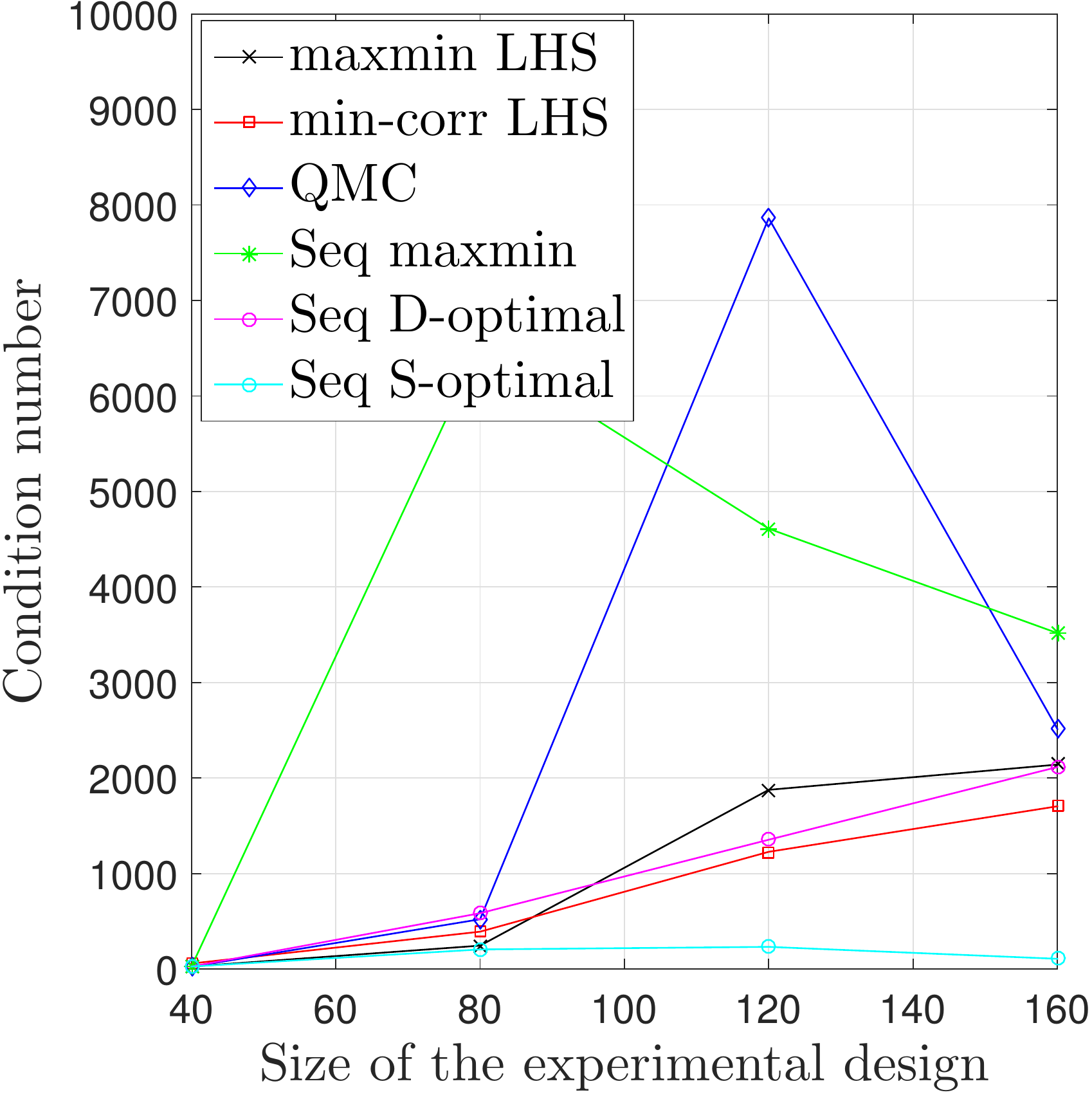}
	
	\caption{The average, over 50 trials of the $maximin$ distance, discrepancy measure and the condition number for the 5 tested experimental designs for the Ishigami function. Top left (a): Variation of $maximin$ distance with increasing size of the ED; Top right (b): Variation of Discrepancy measure with increasing size of the ED; Bottom (c): Variation of condition number criterion with increasing size of the ED;}
	\label{fig:1}
\end{figure}

A sparse, degree-adaptive PCE approach with maximum degree in the range 3-20 is chosen. The PCE coefficients are computed with the hybrid LAR method \citep{Blatman2010b}. The accuracy of the PCE is assessed with a MCS-based validation set of size $N_{val}=10^6$. The sequential sampling methods ($Seq$ S-optimal, $Seq$ D-optimal and $Seq$ maximin) are initialized with an experimental design generated by maximin LHS of size $40$, iteratively augmented with $N_{add}=20$ additional points. For the other sampling methods ($maximin$ LHS, $min-corr$ LHS), a predefined size is considered. As already mentioned, the analysis is replicated 50 times to assess the statistical uncertainty. The performance  of the different sampling methods is compared in terms of their RMSE on the validation set for varying ED sizes. \\

As a first step, the performances of the various design strategies with respect to the space-filling quality measures introduced in Section \ref{sec:measures of quality} are compared. The average values of the different measures calculated from the $50$ repetitions for various sizes of the ED are plotted in Figure ~\ref{fig:1}. The top left panel of Figure \ref{fig:1} reports results for the maximin distance $D_{min}$ criterion. As expected, the sequential $maximin$ experimental design systematically shows the maximal minimum distance across all designs.  All the other designs show a consistent decrease in the minimal distance with increasing ED size, leading to a deterioration of their space filling properties. The worst results are obtained with D-optimal design.

In the same format, the top-right panel of Figure~\ref{fig:1} reports results for the $discrepancy$ criterion. All the non-sequential strategies tested (LHS and QMC) show an approximately constant 
discrepancy regardless of the experimental design size. However, all the sequential strategies show instead a constant increase with the experimental design size. This means that all the sequential strategies tend to produce samples that deviate from the uniform distribution. In other words, design 
strategies that are optimal for regression problems do not necessarily exhibit a
low-discrepancy behaviour.

The condition number of the model matrices are plotted with respect to the size of the ED in the lower panel of Figure~\ref{fig:1}. As expected, $Seq$ S-optimal designs show a consistently smaller condition number for all experimental designs sizes. $Seq$ D-optimal, $maximin$ LHS and $min-corr$ LHS exhibit comparable behaviours with a monotone increase in condition number. The remaining $Seq$ maximin ED and Sobol' sequences show instead a somewhat more erratic behaviour. Overall, Figure~\ref{fig:1} shows that $Seq$ S-optimal designs provides an overall good balance between the three criteria.

\begin{figure}
	\centering
	\includegraphics[width=.4\textwidth,clip = true, trim = 0 0 0 
	0]{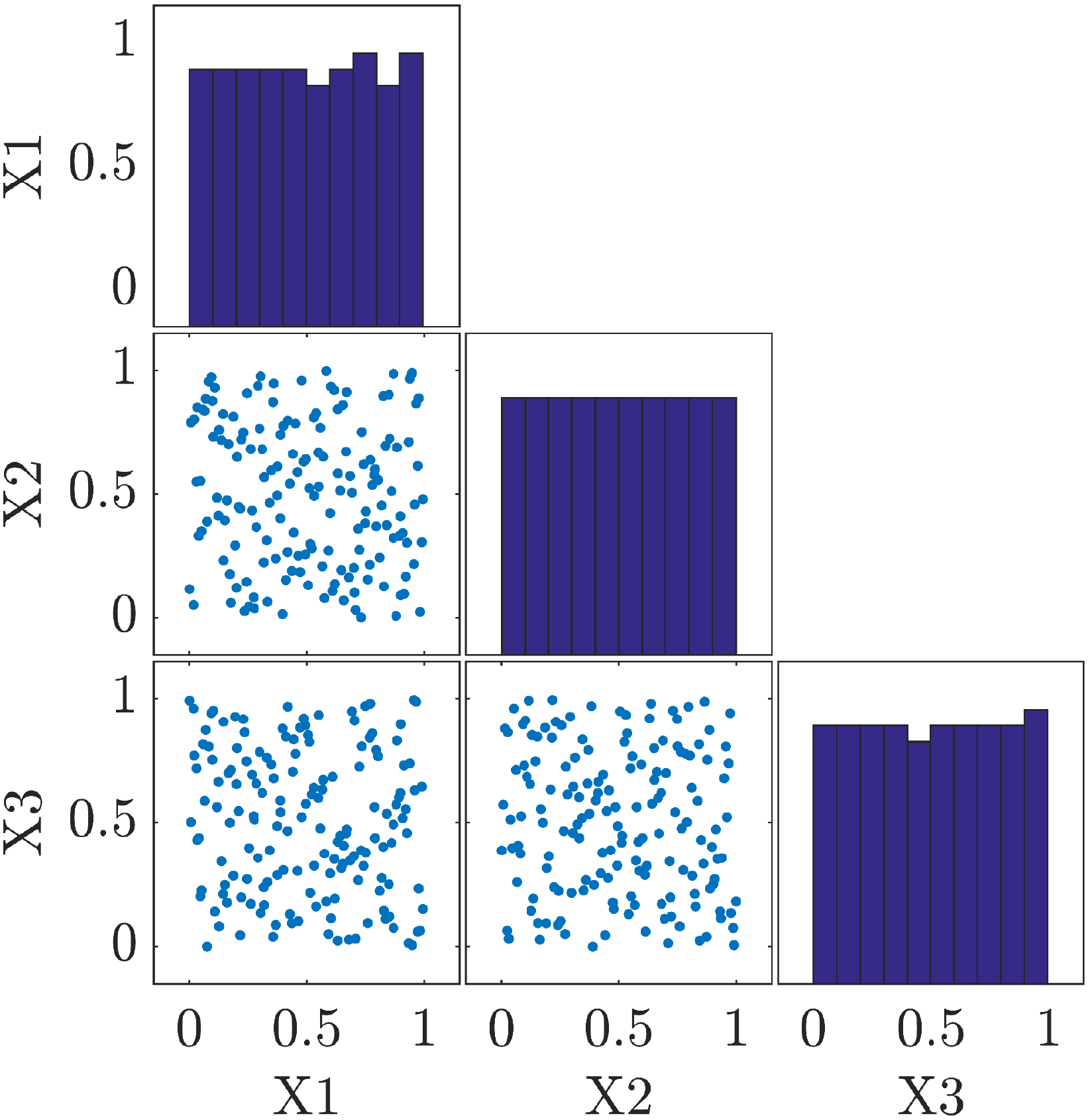}
	\includegraphics[width=.4\textwidth,clip = true, trim = 0 0 0 
	0]{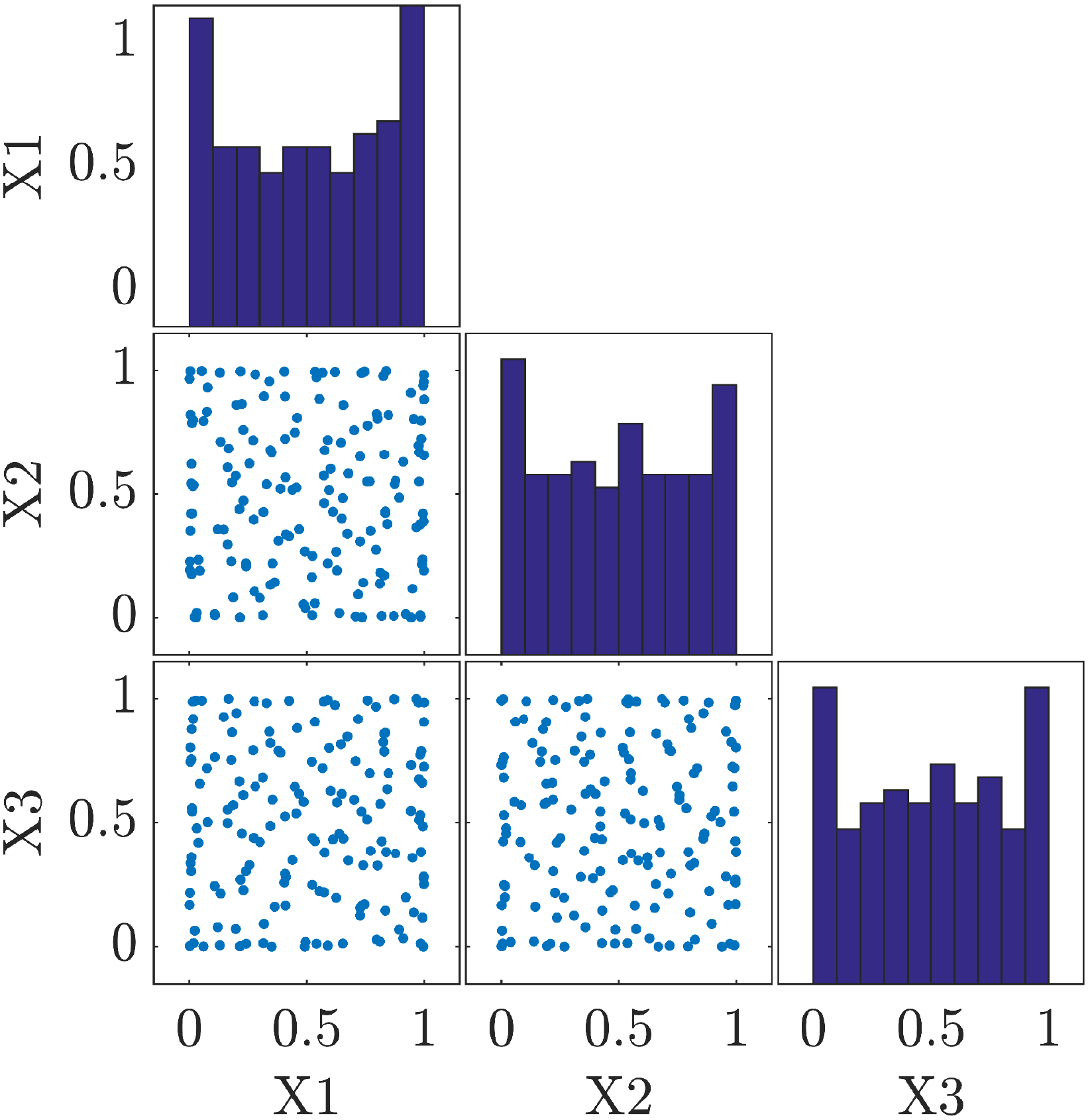}
	\includegraphics[width=.4\textwidth,clip = true, trim = 0 0 0 
	0]{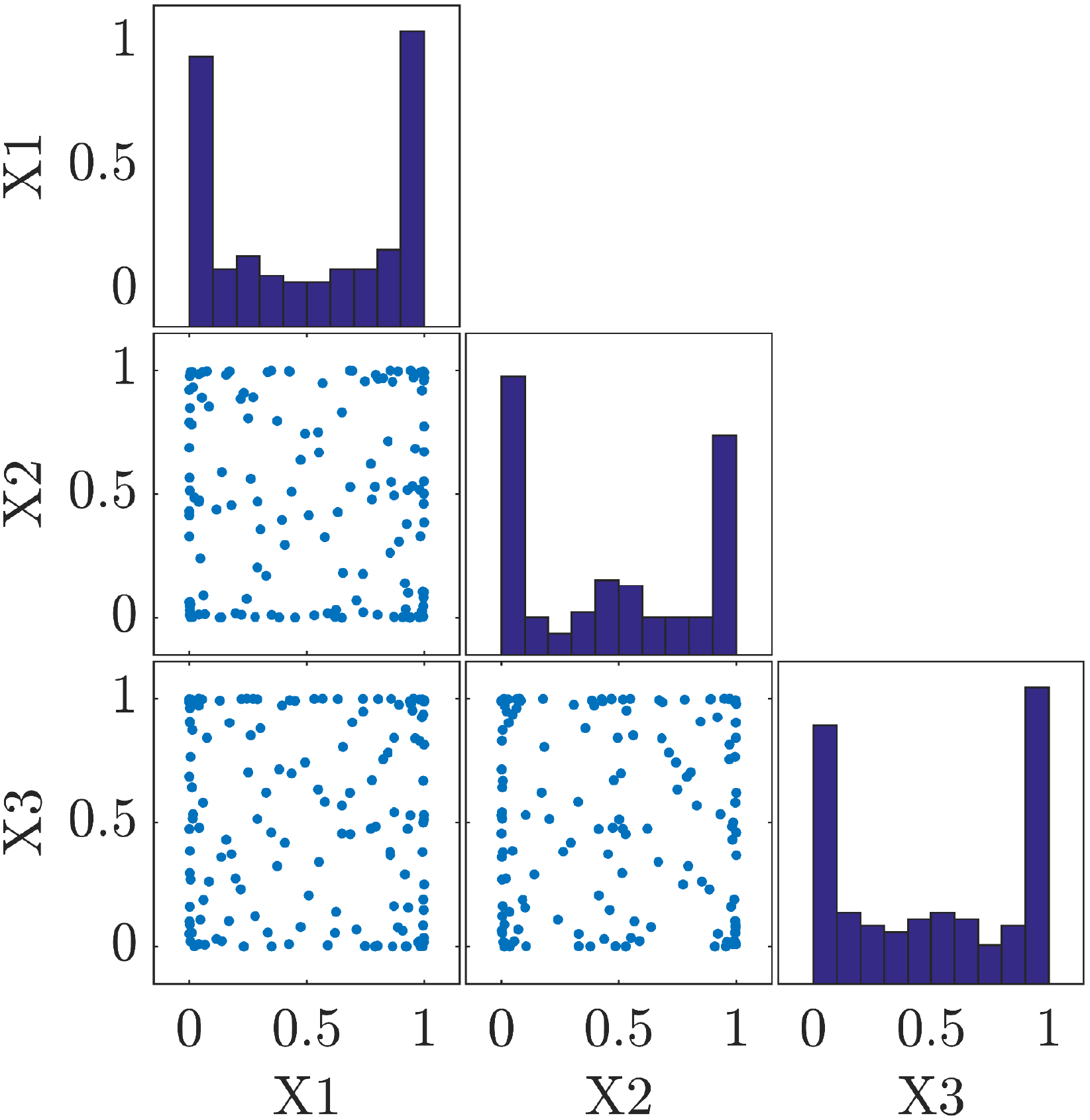}
	\includegraphics[width=.4\textwidth,clip = true, trim = 0 0 0 
	0]{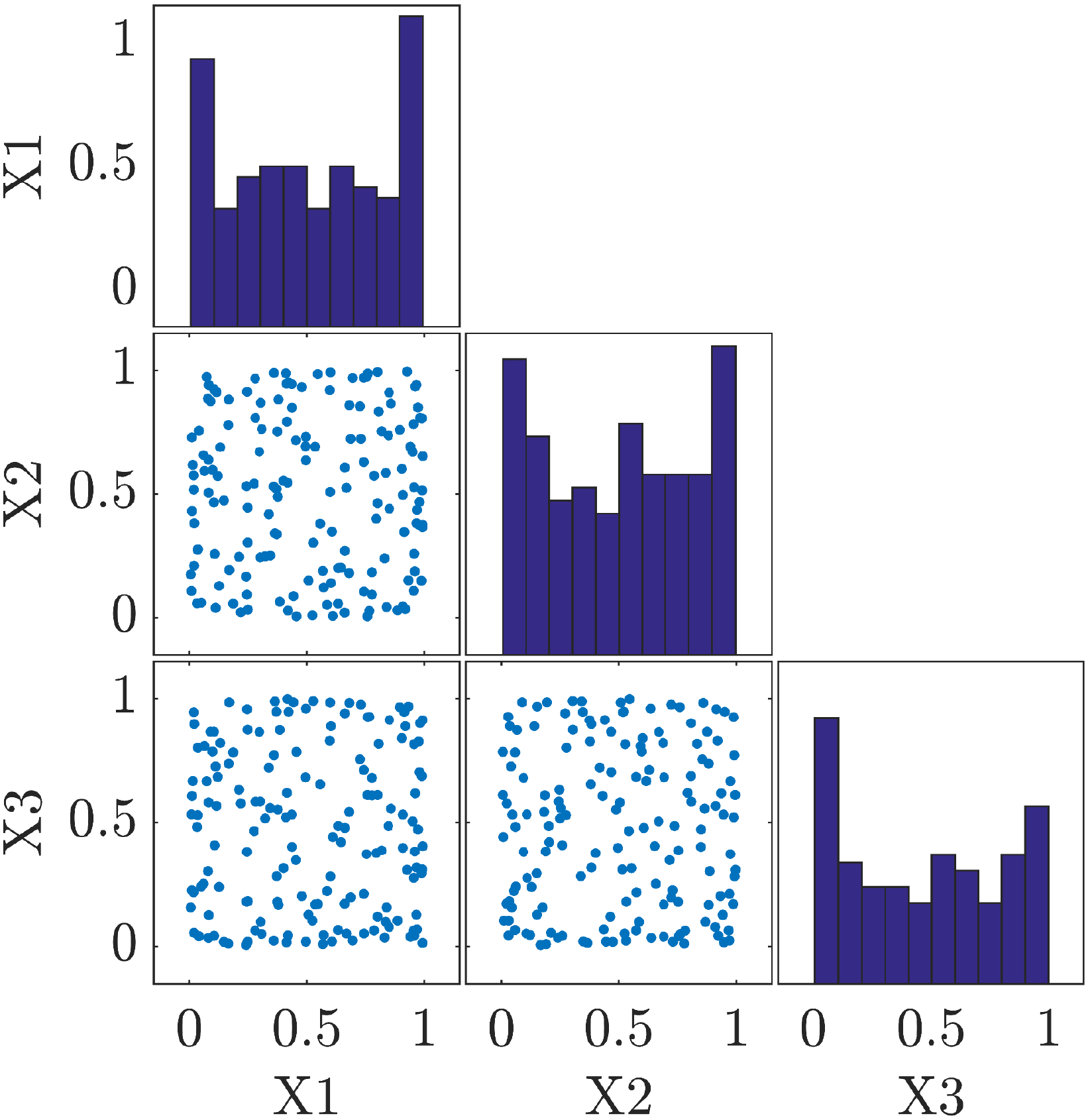}
	\caption{An example of 2D plots of a 3-dimensional designs of size 160. Top left: Experimental design obtained with maximin LHS;  Top right: Experimental design obtained with Seq maximin; Bottom left: Experimental design obtained with maximin sequential D-optimal;  Bottom right: Experimental design obtained with sequential $S$-optimal; }
	\label{fig:2}
\end{figure}

To gain an intuitive understanding of the different 
experimental designs, Figure ~\ref{fig:2} 
compares 2D scatter plots and histograms for the final ED of size $N=160$ generated by 
$maximin$ LHS and the three sequential designs $Seq$ maximin, $Seq$ D-optimal and $Seq$ S-optimal. Uniform  
space coverage is obtained onlywith $maximin$ LHS. The sequential design 
obtained with $Seq$ maximin and $Seq$ S-optimal designs appear to be 
fairly uniformly distributed, with some degree of 
accumulation on the boundaries. On the contrary, $Seq$ D-optimal sample points tend to cluster near the boundaries of the design space. \\

\begin{figure}[h]	
	\centering
	{\includegraphics[width=.5\textwidth,clip = true, trim = 0 0 0 
		0]{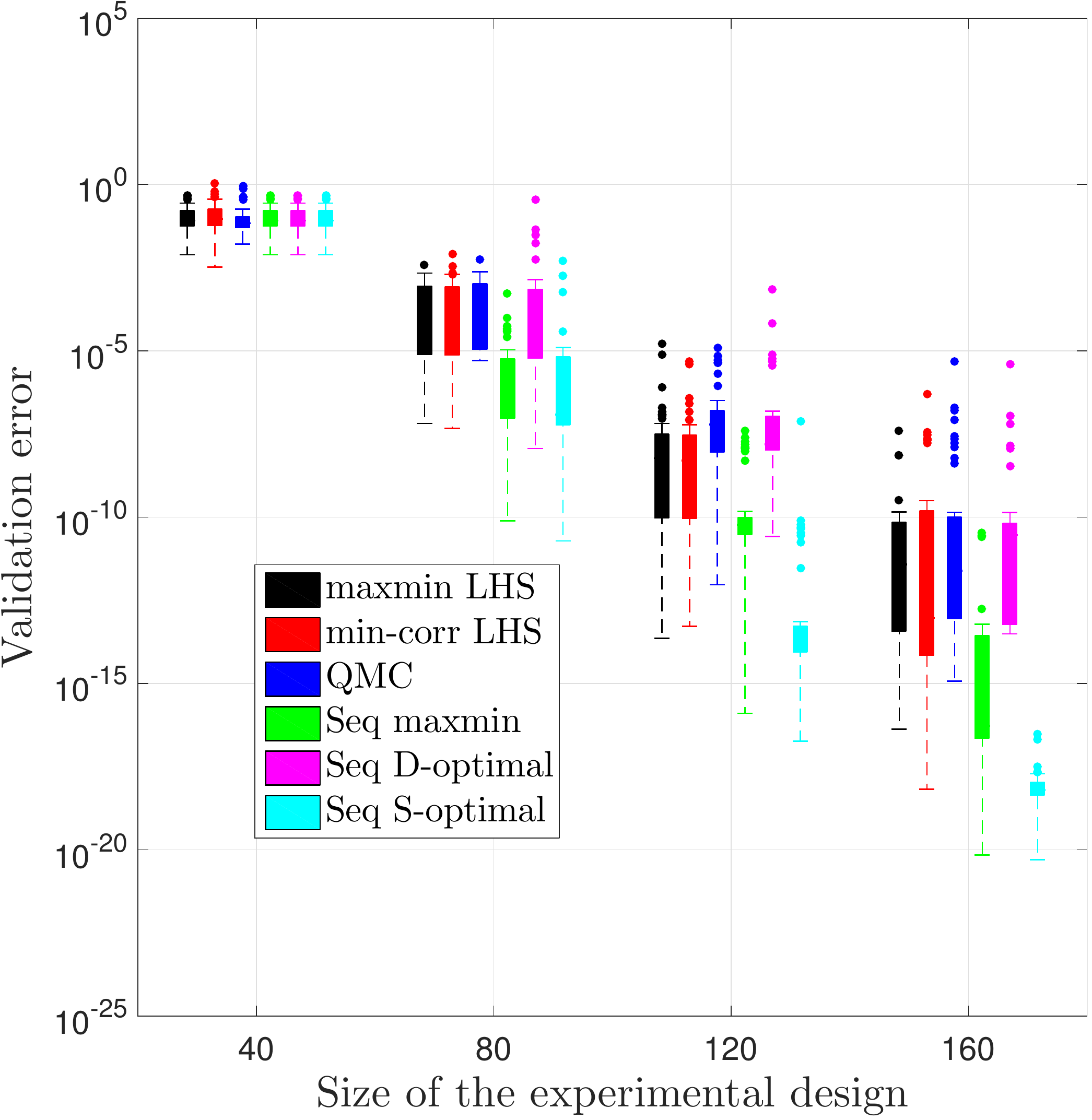}}
	\caption{Evolution of boxplots of the validation error for varying
		sizes of the experimental design for the Ishigami function. The boxplots are obtained from 50 replications.}
	\label{fig:3}
\end{figure}

The metamodelling performance of the different sampling strategies are compared in Figure ~\ref{fig:3}. The boxplots represent the median RMSE error, the $25$\% and $75$\% percentiles and the extreme error values (outliers) of the 50 independent runs for varying ED sizes, ranging from $N = 40$ to $N = 160$. We observe that $Seq$ S-optimal design consistently outperforms other sampling strategies. In addition, it generally shows a more stable behaviour with smaller variability between repetitions, especially as the size of the experimental design increases. Accurate results are also achieved with $Seq$ maximin design. The prediction improvement seems to be more important for larger experimental designs. We note that comparable results are obtained with $maximin$ LHS, $min-corr$ LHS and Sobol' sequences.

\subsection{Sobol function}
Let us consider now the so called Sobol' g-function: 
\begin{equation} 
\label{eqn:28}
f(\ve{x}) = \prod_{i=1}^q {\frac{ | 4x_i-2| +a_i}{1+a_i}}
\end{equation}
The input parameters $\{X_i, i=1,...,q\}$ are \textit{i.i.d. uniform} distributions over $[0, 1]$. 
The $\{a_i, i=1,...,q\}$ are parameters controlling the complexity of the function. For our numerical application, we choose to fix $q=8$ and $a=[1, 2, 5, 10, 20, 50, 100, 500]$. The same 
analysis presented for the Ishigami function was conducted to illustrate the performance of the proposed approach for this benchmark. The Sobol' function is highly complex (strongly nonlinear, non-monotonic and non-polynomial), therefore requiring high polynomial degree even for low dimensions. The truncation options $p = 30$ and $q = 0.5$ are used for the 
PCE calculation. The initial experimental design is generated by 
a maximin LHS of size $100$, augmented iteratively with $N_{add}=50$ additional 
points. \\

\begin{figure}[h]	
	\centering
	{\includegraphics[width=.5\textwidth,clip = true, trim = 0 0 0 
		0]{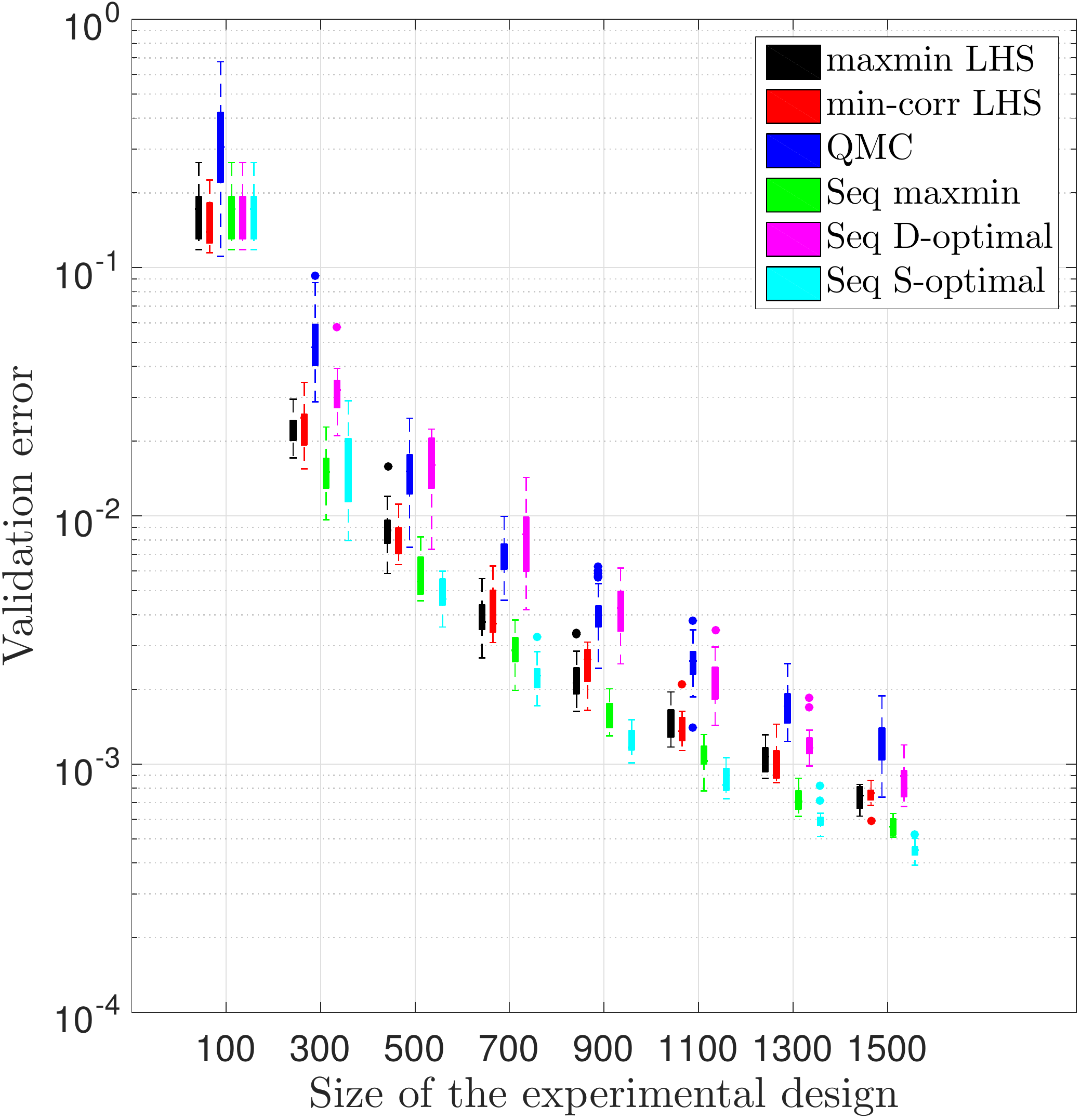}}
	\caption{Evolution of boxplots of the validation error for varying
		sizes of the experimental design for the Sobol function. The boxplots are obtained from 50 replications.}
	\label{fig:5}
\end{figure}

The performances of the various design strategies with respect to the space-filling quality measures have been compared, but they are not shown herein because the general conclusions drawn for the Ishigami function still hold. The metamodelling performance of the different sampling strategies are compared in Figure ~\ref{fig:5}. As in the previous example, the $Seq$ S-optimal design outperforms the others sampling methods (Figure ~\ref{fig:5}). Good performance are also achieved with $Seq$ maximin design as in the Ishigami function.  Furthermore, much smaller variances (boxplots are smaller) are shown for $Seq$ S-optimal design and lead to the conclusion that this class of designs is robust. Sobol' sequences give the least accurate results. This behaviour may be due to the increase in the dimensionality of the problem, coupled with the tendancy of Sobol' sequences to cluster on regular sub-manifolds in high dimension \citep{morokoff1994quasi}. 

\subsection{Maximum deflection of a truss structure}
\begin{figure}[h]
	\centering
	\includegraphics[width=.45\textwidth,clip = true, trim = 0 0 0 
	0]{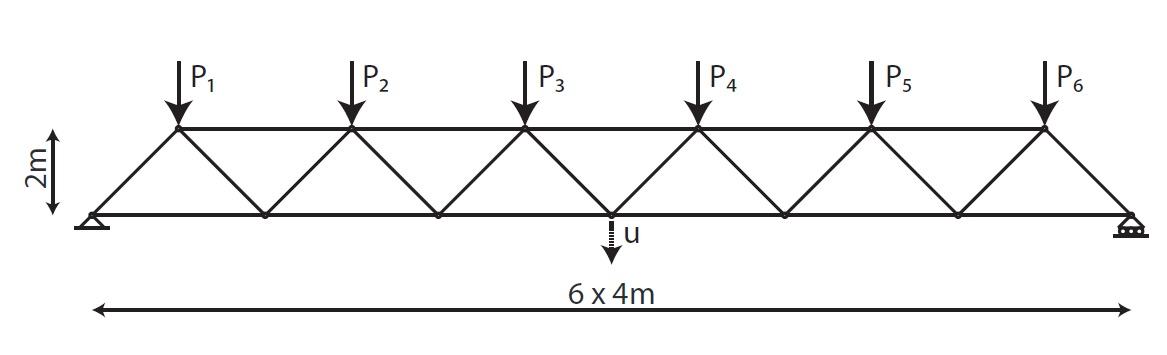}
	\caption{Truss structure with 23 members.}
	\label{fig:6}
\end{figure}

Let us consider the truss structure shown in Figure ~\ref{fig:6}. Ten independent variables are considered, namely: the vertical loads, $P_1, . . . ,P_6$; the cross-sectional area and Young's modulus of the horizontal bars, $A_1$ and $E_1$, respectively; the cross-sectional area and Young's modulus 
of the vertical bars, $A_2$ and $E_2$, respectively (Figure ~\ref{fig:6}).
The model input parameters are described by independent random variables whose distributions are reported in Table 1. An isoprobabilistic transformation of $\ve{X}$ into standard normal variables is employed. In the underlying deterministic problem, the truss deflection $u$ is computed with a finite-element \textsc{Matlab} code. The same study as in the two previous examples is conducted for this 10-dimensional function.
\begin{table} [h]
	\centering
	\caption{Truss-deflection problem: random input variables.}
	\label{tab:truss_input}
	\begin{tabular}{c c c c}
		\hline Variable & Distribution & Mean & CoV \\
		\hline $A_1~[\rm m]$ & Lognormal & $0.002$ & $0.10$ \\
		$A_2~[\rm m]$ & Lognormal & $0.001$  & $0.10$ \\
		$E_1, E_2~[\rm MPa]$ & Lognormal & $2.1 \cdot 10^5$   & $0.10$\\
		$P_1 \cdots  P_6~[\rm KN]$ & Gumbel & $50$  & $0.15$ \\
		\hline       
	\end{tabular}
\end{table}

A sparse degree-adaptive PCE approach with maximum degree in the range $3-20$ is chosen. The accuracy of the PCE is assessed with a MCS-based validation set of size $N_{val}=10^6$. The sequential sampling methods ($Seq$ S-optimal, $Seq$ D-optimal and $Seq$ maximin) are initialized with an experimental design generated by maximin LHS of size $100$, augmented iteratively with $N_{add}=50$ additional points.  For the other sampling methods ($maximin$ LHS, $min-corr$ LHS), a predefined size is considered. The performance  of the different sampling methods is compared in terms of their RMSE on the validation set for varying ED sizes. \\

\begin{figure}[h]	
	\centering
	{\includegraphics[width=.46\textwidth,clip = true, trim = 0 0 0 
		0]{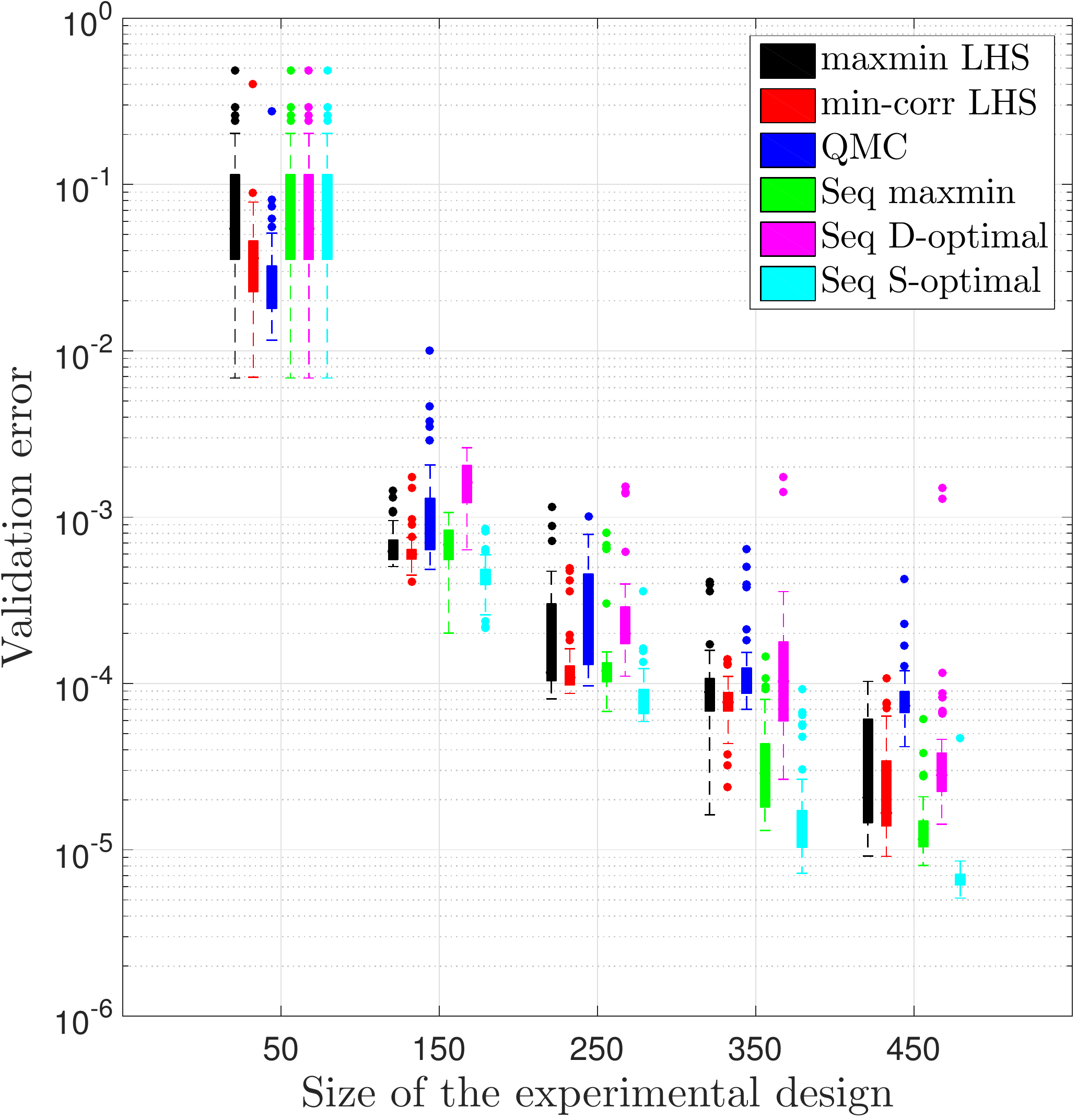}}
	\caption{Evolution of boxplots of the validation error for varying
		sizes of the experimental design for the truss function. }
	\label{fig:8}
\end{figure}

As in the previous cases, the performances of the various design strategies with respect to the space-filling quality measures were compared but the results are not shown though for the sake of conciseness, as they are in line with the previous two cases. The metamodelling performance of the different sampling strategies are compared in Figure ~\ref{fig:8}. As in the two previous examples, the $Seq$ S-optimal design performs better than the other sampling methods, with a faster decrease of the validation errors. One sees that the computational budget gain increases exponentially with the size of the ED. Fairly accurate results are also obtained with $Seq$ maximin design. Similar accuracies are obtained with $maximin$, $min-corr$ LHS. Again, a poor performance is observed for Sobol' sequences and $D$-optimal designs, especially for large experimental designs.

\subsection{One-dimensional diffusion problem}

In this section, we consider a standard benchmark in UQ computations \citep{shin2016nonadaptive}, a one-dimensional stochastic diffusion problem subject to uncertainty in the diffusion coefficient. Let us consider the following boundary-value problem defined over a domain 
$\mathcal{D} = [0 ; L]$:
\begin{equation} 
\label{eqn:29 }
\begin{split}
-\frac{d}{d x}\left[E(x,\omega)\frac{d u}{d x}(x,\omega)\right]+f(x)=0\\
u(0)=0\\
\frac{du}{dx}(L)=F.
\end{split}
\end{equation}
Furthermore, assume that the diffusion coefficient is a 
log-normal random field described by: 
\begin{equation} 
\label{eqn:30}
E(x,\omega)=\exp(\lambda_E+\zeta_E g(x,\omega)),
\end{equation}
where $g(x,\omega))$ is a standard normal stationary Gaussian random field with an exponential autocorrelation function:
\begin{equation} 
\label{eqn:31}
\rho(x,x')=e^{-|x'-x|/l}.
\end{equation}
The Gaussian random field $g(x)$ in ~\eqref{eqn:30} is approximated using the Karhunen-Lo\`eve Expansion \citep{loeve1977probability}:
\begin{equation} 
\label{eqn:32}
g(x,\omega)=\sum_{k=1}^{M} \sqrt{\iota_k}  \varphi_k(x)\xi_k(\omega),
\end{equation}
Where $\{(\iota_k;\varphi_k), k=1,...,M \}$ are the solutions of the 
Fredholm equation associated to the exponential kernel, see 
\citet{Ghanembook1991,Sudret2000} for the detailed analytical solution.\\

For the numerical test we set the following numerical values: $L = 1$, $F = 
1$, $f = 0.5$. We set: $\lambda_E=10 $, $\zeta_E=3 $ and $l=1/3$. The number of 
terms to be retained in \eqref{eqn:32} is selected such that: 
\begin{equation} 
\label{eqn:33}
\sum_{k=1}^{M} {\iota_k}/ \sum_{k=1}^{\infty} {\iota_k} \geq 0.99.
\end{equation}
$M = 62$ are therefore required, making this the highest dimensional example considered.. The quantity of interest (model response) is $u(L)$.\\

A sparse, degree-adaptive PCE approach with maximum degree in the range $1$-$4$ is chosen with $q=0.5$. It is worth noting that the model is extremely sparse with a low effective dimension, although its high dimensionality. For the sequential sampling methods, the following conditions are then set: an initial experimental design of size $N_{init}= 100$ generated with $maximin$ LHS, $N_{add}=20$ are considered. \\

\begin{figure}[h]	
	\centering
	{\includegraphics[width=.5\textwidth,clip = true, trim = 0 0 0 
		0]{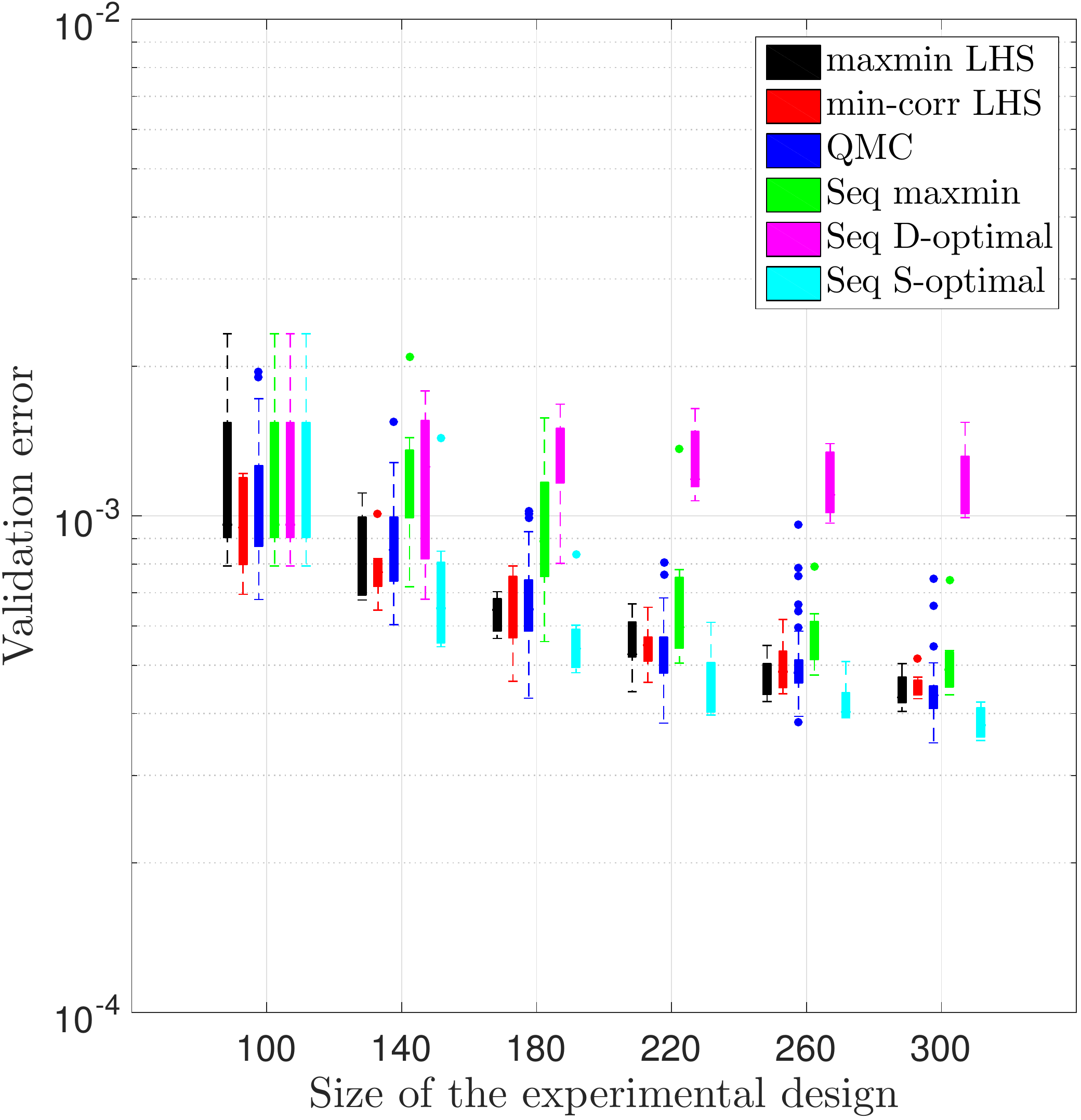}}
	\caption{Evolution of boxplots of the validation error for varying
		sizes of the experimental design for the 1D diffusion function. }
	\label{fig:10}
\end{figure}

We now investigate the metamodelling performance of the different sampling strategies. Figure ~\ref{fig:10} shows boxplots of the RMSE error. As in the previous examples, the $Seq$ S-optimal performs better than the other sampling methods. Unlike the previous examples, $Seq$ maximin designs do not perform better than $maximin$ LHS and $min-corr$ LHS. It is worth to note that $Seq$ maximin designs search the design in $62$ dimensions and does not capture the sparsity of the model (i.e.; the model has a few variables that are significantly more influential than the others). Again, a poor performance is observed for the Seq $D$-optimal designs.

\subsection{Effects of the choice of $N_{init}$ and $N_{Add}$}
\label{sec:N_ini vs N_add}

\begin{figure}
	\centering
	\includegraphics[width=.45\textwidth,clip = true, trim = 0 0 0 
	0]{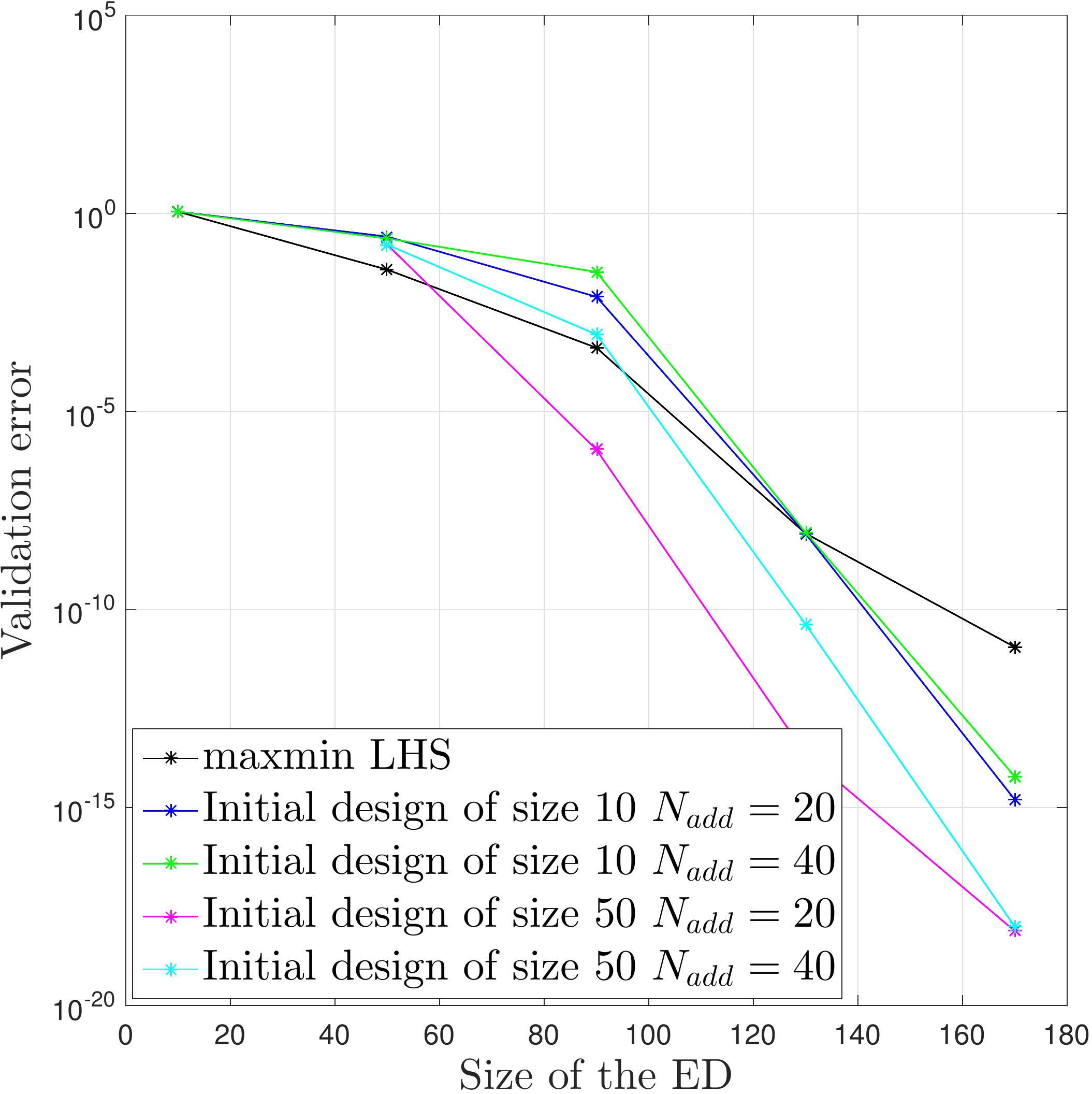}
	\includegraphics[width=.45\textwidth,clip = true, trim = 0 0 0 
	0]{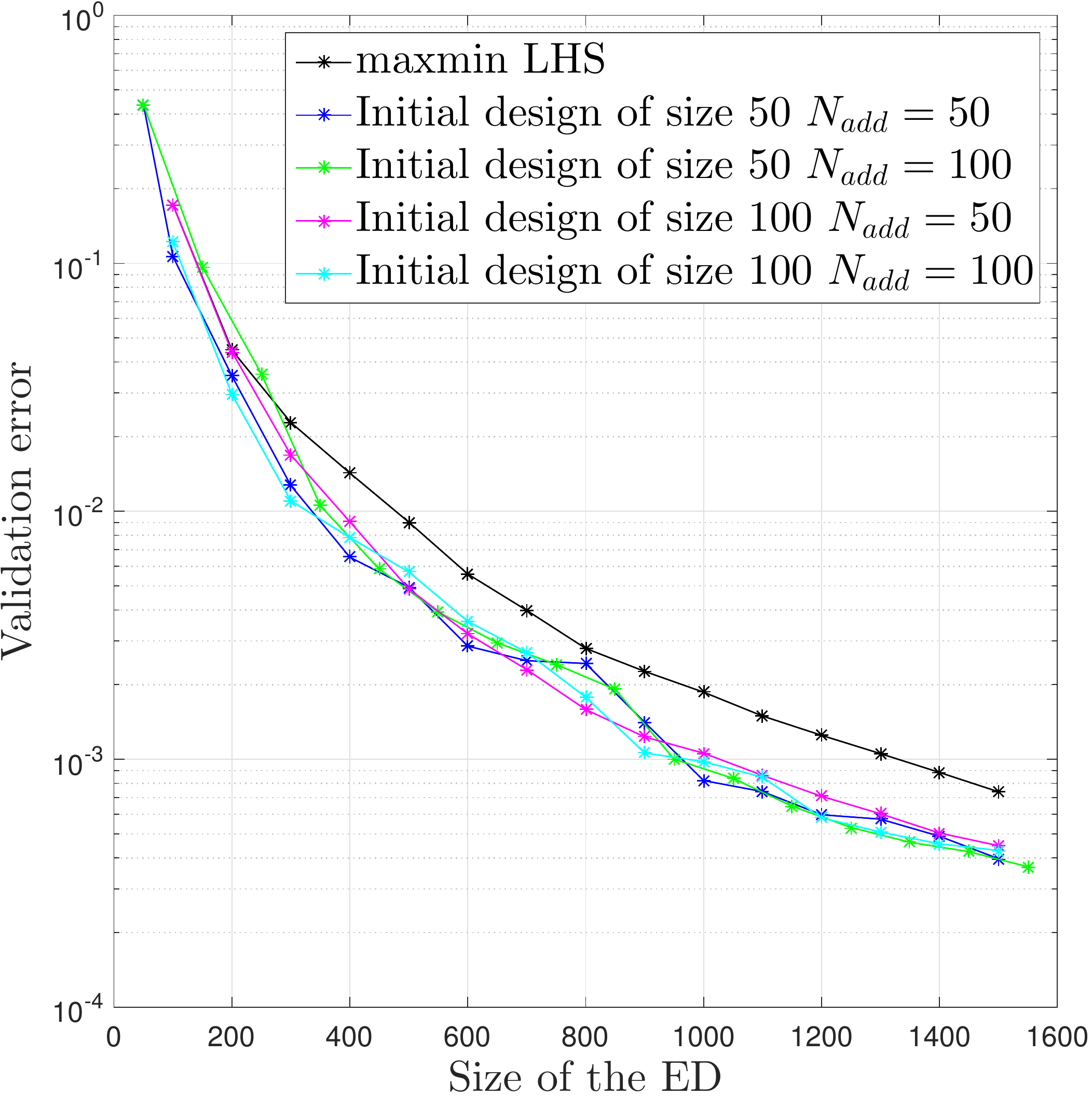}
	\includegraphics[width=.45\textwidth,clip = true, trim = 0 0 0 
	0]{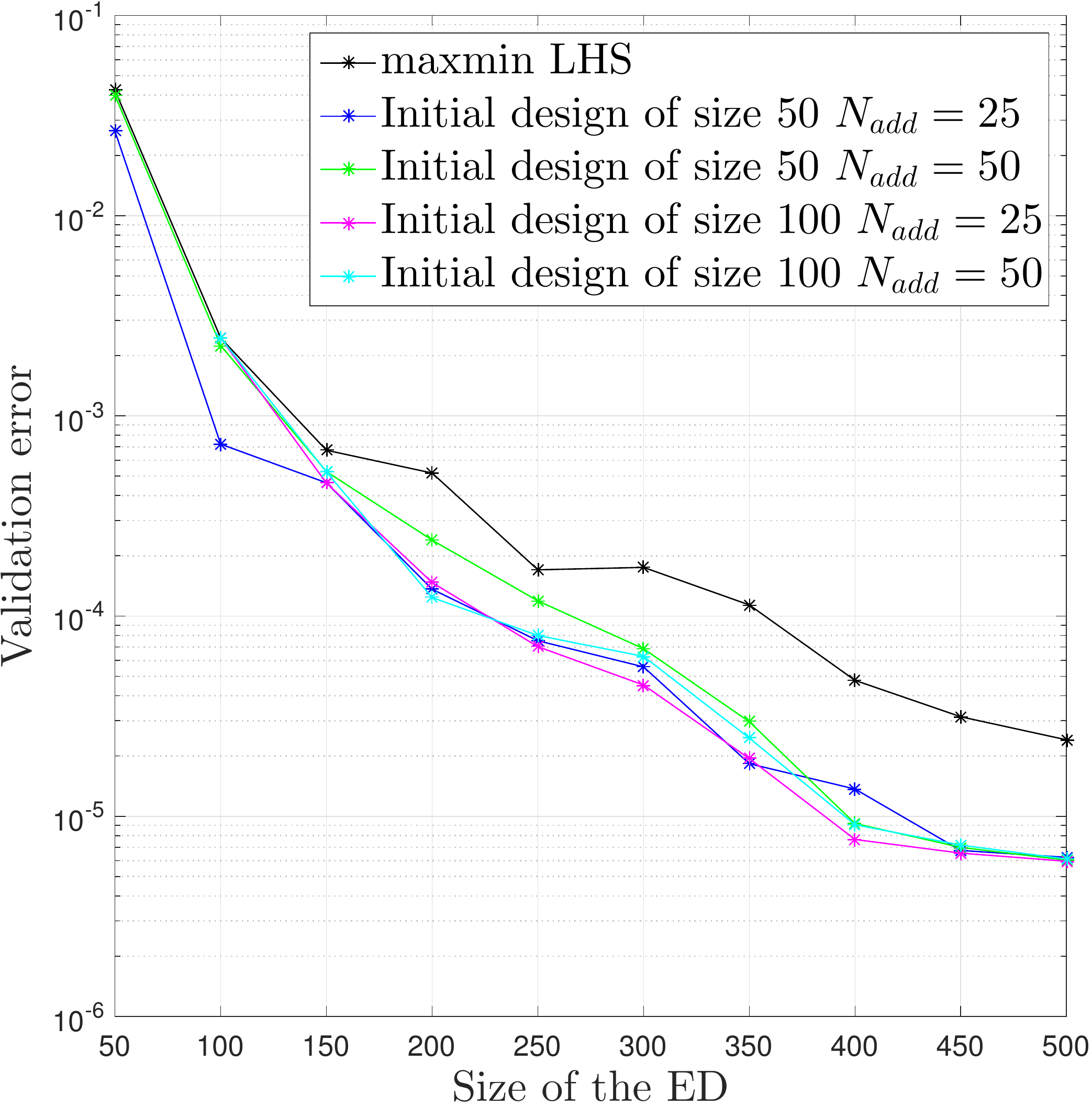}
	\includegraphics[width=.45\textwidth,clip = true, trim = 0 0 0 
	0]{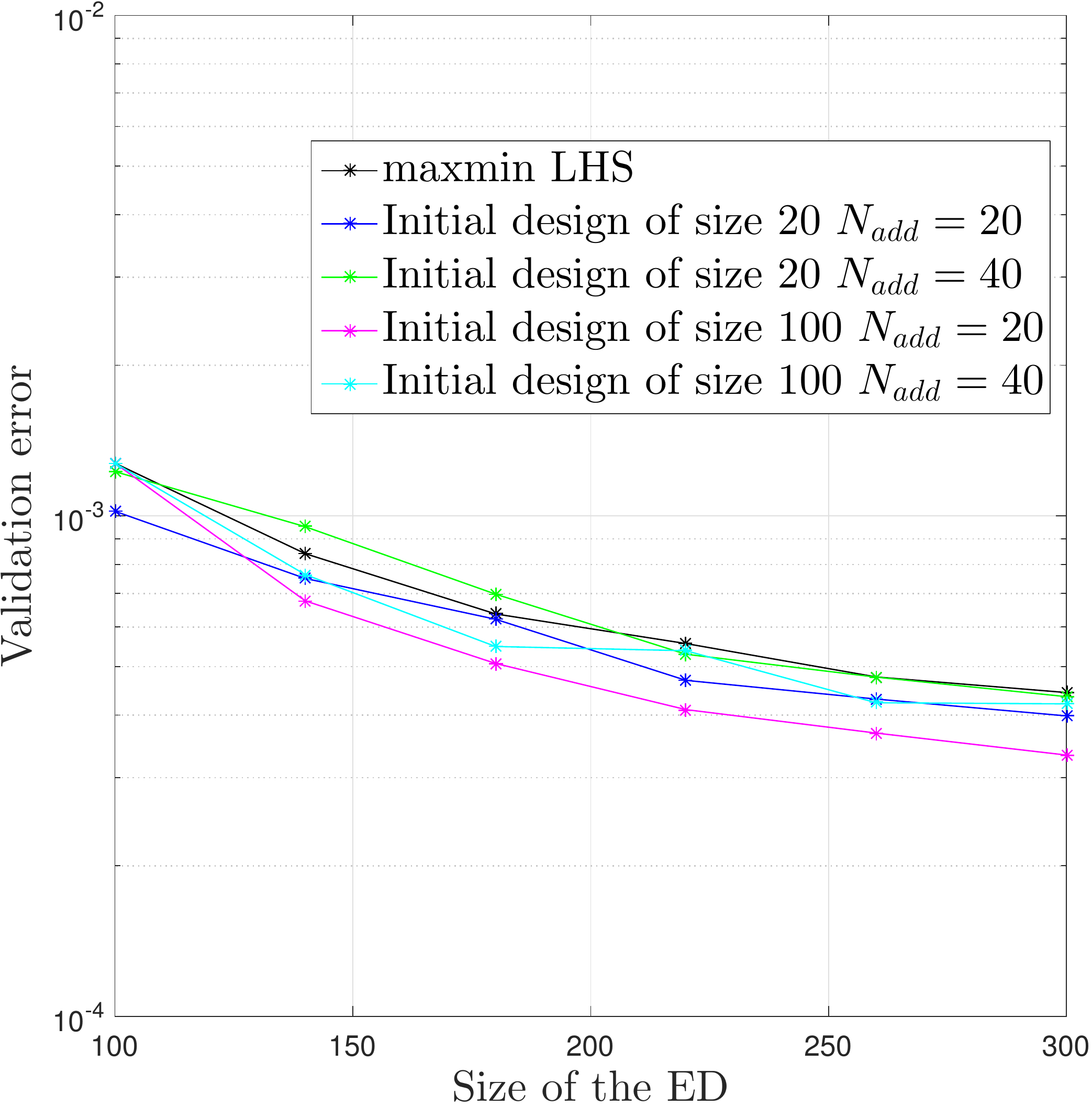}
	\caption{Effect of the size of the initial design and $N_{add}$ points. Top left: Ishigami function; Top right: Sobol function; Bottom left: Truss function; Bottom right: 1D diffusion problem}
	\label{fig:11}
\end{figure}

Determining the optimal sample size of the initial design in the sequential construction is a difficult task. In the context of sequential sampling, previous literature does not have a well-defined rule of thumb for the number of initial runs, given a total budget $N$. According to \citet{bernardo1992integrated}, the initial design should at least contain three observations per input variable. \citet{jones1998efficient} recommended larger initial designs with up to ten observations per input variable. A similar question holds on the choice of $N_{add}$. It is likely that the optimal $N_{init}$ and $N_{add}$  depend on the nature of the model that is to be approximated and on its degree of sparsity. \\
To provide some insight into the effects of those choices on the $Seq$ $S$-optimal design class, a parametric study was performed on the examples presented in the previous sections. For each of the examples provided, several choices of $N_{init}$ and $N_{add}$ were compared in terms of the final surrogate RMSE prediction. The most significant results are reported in Figure~\ref{fig:11} for $Seq$ $S$-optimal designs.
The results for the Ishigami function suggest that indeed the choice of $N_{init}$ and $N_{add}$ plays a significant role in the accuracy of PCE for all of the considered ED sizes, with best performances achieved with a relatively large initial experimental design. However, no comparable trends were observed for either of the remaining benchmarks in the other panels of Figure~\ref{fig:11}. 

These results show that no reliable rule of thumb can be determined to choose $N_{init}$ and $N_{add}$. The idea is to start with a relatively small sample to obtain initial information about the global behaviour of the metamodel. This information will be updated through iterations. However, if the initial design is too 
small, the obtained information is not sufficient to achieve an accurate estimate of the optimal basis from LARS, leading to a poor optimization of the $S$-value criterion in the early iterations. A rather large initial design will not ameliorate the results despite an increased computational effort.

\section{Conclusions}

The paper handles the topic of selecting the best design of experiments in order to ensure the optimal accuracy of sparse polynomial chaos expansions over the whole input space within a given computational budget. The coefficients of the PCE are computed based on least-square analysis by relying on the so-called experimental design. An efficient approach for constructing an optimal experimental design for sparse polynomial chaos expansions was proposed. The complete approach includes two major elements: (a) the use of degree-adaptive sparse PCE based on LARS for selecting the optimal basis, and (b) the choice of an efficient criterion for selecting the optimal sample points. The ED is build up sequentially by enriching an initial design, which is generated using a space-filling technique. The added points are chosen according to an optimality criterion that depends on the current sparse basis. In this work, two optimality criteria are used, that is: (i) D-optimality and (ii) S-value. The D-optimal criterion is related to the determinant of the information matrix whereas the S-value criterion is related to the orthogonality of the columns of the information matrix.

A comprehensive comparative study of the different sampling strategies was performed on four numerical models with varying input dimensionality and complexity. To complete our investigation, a sequential strategy based on optimizing the well-known maximin distance criterion was also studied. Our results lead to the following major conclusions.
\begin{itemize}
	\item[$\bullet$] The use  of a sequential adaptive design according to the $S-$value criterion, $Seq$ S-optimal, ensures optimal performances in terms of the relative generalization error. The proposed approach  provides a significant accuracy and stability improvement compared to other sampling strategies.
	\item[$\bullet$] The $Seq$ D-optimal approach has shown poor performances.
	\item[$\bullet$] The sequential maximin designs performed well when the number of random input parameters is low to moderate, say less than 10. However, an erratic behaviour is observed for high dimensional cases. 
\end{itemize}

Finally,  further work may include using other basis selection approaches such as Orthogonal Matching Pursuit (OMP) \citep{ Jakeman2015,mallat1993matching} as well as metamodelling techniques such as Low Rank Approximations (LRA) \citep{chevreuil2015least,Konakli201664,konakli2016polynomial,konakli2016reliability} to identify the optimal ED. Future research may also investigate the use of the sequential adaptive designs in the framework of reliability analysis.












\bibliographystyle{chicago}
\bibliography{bibIJUQ}
\end{document}